 \documentclass[review,authoryear,3p,times]{elsarticle}

\usepackage{multirow,setspace,amssymb,amsmath,graphicx,color,rotating,subfigure,url,lineno,natbib}
\usepackage{textcomp}
\usepackage[nofiglist,notablist]{endfloat}

\journal{Journal of Economic Behavior \& Organization} 

\begin{document}

\begin{frontmatter}

\title{Bubble Diagnosis and Prediction \\of the 2005-2007 and 2008-2009
Chinese stock market bubbles}
\author[BS,SS,RCE]{Zhi-Qiang~Jiang}
\author[BS,SS,RCE,RCFE]{Wei-Xing~Zhou}
\author[ETH,UG]{Didier~Sornette~\corref{cor}}
\ead{dsornette@ethz.ch}
\author[ETH]{Ryan~Woodard}
\author[FOR]{Ken~Bastiaensen~\corref{fortisdisclaimer}}
\author[FOR]{Peter~Cauwels~\corref{fortisdisclaimer}}

\cortext[fortisdisclaimer]{These authors express their personal views, which do
  not necessarily correspond to those of BNP Paribas Fortis.}

\cortext[cor]{Corresponding author. Address: KPL F 38.2, Kreuzplatz
5, Chair of Entrepreneurial Risks, Department of Management,
Technology and Economics, ETH Zurich, Switzerland, Phone: +41 44 632
89 17, Fax: +41 44 632 19 14.}

\address[BS]{School of Business, East China University of Science and
  Technology, Shanghai 200237, China}
\address[ETH]{The Financial Crisis Observatory \\Department of Management, Technology and Economics, ETH Zurich,
  Kreuzplatz 5, CH-8032 Zurich, Switzerland}
\address[SS]{School of Science, East China University of Science and
  Technology, Shanghai 200237, China}
\address[RCE]{Research Center for Econophysics, East China University of
  Science and Technology, Shanghai 200237, China}
\address[UG]{Swiss Finance Institute, c/o University of Geneva, 40 blvd. Du
  Pont d'Arve, CH 1211 Geneva 4, Switzerland}
\address[RCFE]{Research Center on Fictitious Economics \& Data Science, Chinese
  Academy of Sciences, Beijing 100080, China}
\address[FOR]{BNP Paribas Fortis, Warandeberg 3, 1000 Brussels, Belgium}

\begin{abstract}
  By combining (i) the economic theory of rational expectation bubbles, (ii)
  behavioral finance on imitation and herding of investors and traders and
  (iii) the mathematical and statistical physics of bifurcations and phase
  transitions, the log-periodic power law (LPPL) model has been developed as a
  flexible tool to detect bubbles. The LPPL model considers the
  faster-than-exponential (power law with finite-time singularity) increase in
  asset prices decorated by accelerating oscillations as the main diagnostic of
  bubbles. It embodies a positive feedback loop of higher return anticipations
  competing with negative feedback spirals of crash expectations. We use the
  LPPL model in one of its incarnations to analyze two bubbles and subsequent
  market crashes in two important indexes in the Chinese stock markets between
  May 2005 and July 2009.  Both the Shanghai Stock Exchange Composite index (US
  ticker symbol SSEC) and Shenzhen Stock Exchange Component index (SZSC)
  exhibited such behavior in two distinct time periods: 1) from mid-2005,
  bursting in October 2007 and 2) from November 2008, bursting in the beginning
  of August 2009.  We successfully predicted time windows for both crashes in
  advance \citep{Sornette-2007,
    Bastiaensen-Cauwels-Sornette-Woodard-Zhou-2009-XXX} with the same methods
  used to successfully predict the peak in mid-2006 of the US housing bubble
  \citep{Zhou-Sornette-2006b-PA} and the peak in July 2008 of the global oil
  bubble \citep{Sornette-Woodard-Zhou-2009-PA}.  The more recent bubble in the
  Chinese indexes was detected and its end or change of regime was predicted
  independently by two groups with similar results, showing that the model has
  been well-documented and can be replicated by industrial practitioners.  Here
  we present more detailed analysis of the individual Chinese index predictions
  and of the methods used to make and test them.  We complement the detection
  of log-periodic behavior with Lomb spectral analysis of detrended residuals
  and $(H,q)$-derivative of logarithmic indexes for both bubbles.  We perform
  unit-root tests on the residuals from the log-periodic power law model to
  confirm the Ornstein-Uhlenbeck property of bounded residuals, in agreement
  with the consistent model of `explosive' financial bubbles
  \citep{Lin-Ren-Sornette-2009-XXX}.
\end{abstract}


\end{frontmatter}

\section{Conceptual framework and the two Chinese bubbles
of 2005-2007 and 2008-2009}
\label{Sec:LPPL:Introduction}

The present paper contributes to the literature on financial bubbles by
presenting two case studies and new empirical tests, in support of the proposal
that (i) the presence of a bubble can be diagnosed quantitatively before its
demise and (ii) the end of the bubble has a degree of predictability.

These two claims are highly contentious and collide against a large consensus
both in the academic literature \citep{Rosser-2008-ACS} and among professionals. For
instance, in his recent review of the financial economic literature on bubbles,
\citet{Gurkaynak-2008-JES} reports that ``for each paper that finds evidence
of bubbles, there is another one that fits the data equally well without
allowing for a bubble. We are still unable to distinguish bubbles from
time-varying or regime-switching fundamentals, while many small sample
econometrics problems of bubble tests remain unresolved.''  Similarly, the
following statement by former Federal Reserve chairman Alan
\citet{Greenspan-2002}, at a
summer conference in August 2002 organized by the Fed to try to understand the
cause of the ITC bubble and its subsequent crash in 2000 and 2001, summarizes
well the state of the art from the point of view of practitioners: ``We, at the Federal Reserve recognized that, despite our
suspicions, it was very difficult to definitively identify a bubble until after
the fact, that is, when its bursting confirmed its existence. Moreover, it was
far from obvious that bubbles, even if identified early, could be preempted
short of the Central Bank inducing a substantial contraction in economic
activity, the very outcome we would be seeking to avoid.''

To break this stalemate, one of us (DS) with Anders Johansen from 1995 to 2002,
with Wei-Xing Zhou since 2002 (now Professor at ECUST in Shanghai) and, since
2008, with the FCO group at ETH Zurich (\url{www.er.ethz.ch/fco/}) have
developed a series of models and techniques at the boundaries between financial
economics, behavioral finance and statistical physics. Our purpose here is not
to summarize the corresponding papers, which explore many different options,
including rational expectation bubble models with noise traders, agent-based
models of herding traders with Bayesian updates of their beliefs, models with
mixtures of nonlinear trend followers and nonlinear value investors, and so on
(see \citet{Sornette-2003} and references therein until 2002 and
the two recent reviews in \citet{Kaizoji-Sornette-2009-XXX,Sornette-Woodard-2009-XXX} and references
therein). In a nutshell, bubbles are identified as ``super-exponential'' price
processes, punctuated by bursts of negative feedback spirals of crash
expectations. These works have been translated into an operational methodology
to calibrate price time series and diagnose bubbles as they develop.  Many
cases are reported in Chapter 9 of the book \citep{Sornette-2003} and more
recently successful applications have been presented with ex-ante public
announcements posted on the scientific international database \url{arXiv.org}
and then published in the referred literature, which include the diagnostic and
identification of the peak time of the bubble for the UK real-estate bubble in
mid-2004 \citep{Zhou-Sornette-2003a-PA}, the U.S. real-estate bubble in mid-2006
\citep{Zhou-Sornette-2006b-PA}, and the oil price peak in July 2008
\citep{Sornette-Woodard-Zhou-2009-PA}.

\citet{Kindleberger-2000} and \citet{Sornette-2003} have
identified the following generic scenario developing in five acts, which is
common to all historical bubbles: displacement, take-off, exuberance, critical
stage and crash. For the Chinese bubble starting in 2005, the ``displacement''
and ``take-off'' can be associated with the split share structure reform of
listed companies in 2005. Before the reform, only about one third of the shares
of any listed company in the Chinese stock market were tradable. The other
two-third shares were non-tradable (not allowed to be exchanged and to
circulate between investors), and were owned by the state and by legal
entities. The tradable stocks acquired therefore a significant liquidity
premium, and were valued much higher than their non-tradable siblings,
notwithstanding the fact that both gave the same privilege to their owners in
terms of voting rights and dividends. Since 2001, the Chinese stock market
entered an anti-bubble phase \citep{Zhou-Sornette-2004a-PA} with the Shanghai
Stock Exchange Composite index falling from its then historical high 2245 on 24
June 2001 to the historical low on June 6, 2005. On 29 April 2005, the China
Securities Regulatory Commission launched the pilot reform of the split share
structure. The split share structure reform is defined as the process to
eliminate the discrepancies in the A-share transfer system via a negotiation
mechanism to balance the interests of non-tradable shareholders and tradable
shareholders. On 4 September 2005, the China Securities Regulatory Commission
enacted the {\em{Administrative Measures on the Split Share Structure Reform of
    Listed Companies}}\footnote{Available at
  \url{http://www.csrc.gov.cn/n575458/n4001948/n4002120/4069846.html}, accessed
  on 30 August 2009.}, which took effect immediately.  It is widely accepted
that the split share structure reform was a turning point which triggered and
catalyzed the recovery of the Chinese stock market from its previous bearish
regime.  For the Chinese bubble starting in November 2008, the ``take-off'' can
be associated with China's policy reaction on the global financial crisis, with
a huge RMB 4 trillion stimulus plan and aggressive loan growth by financial
institutions.

\begin{figure}[htp]
\centering
\includegraphics[width=10cm]{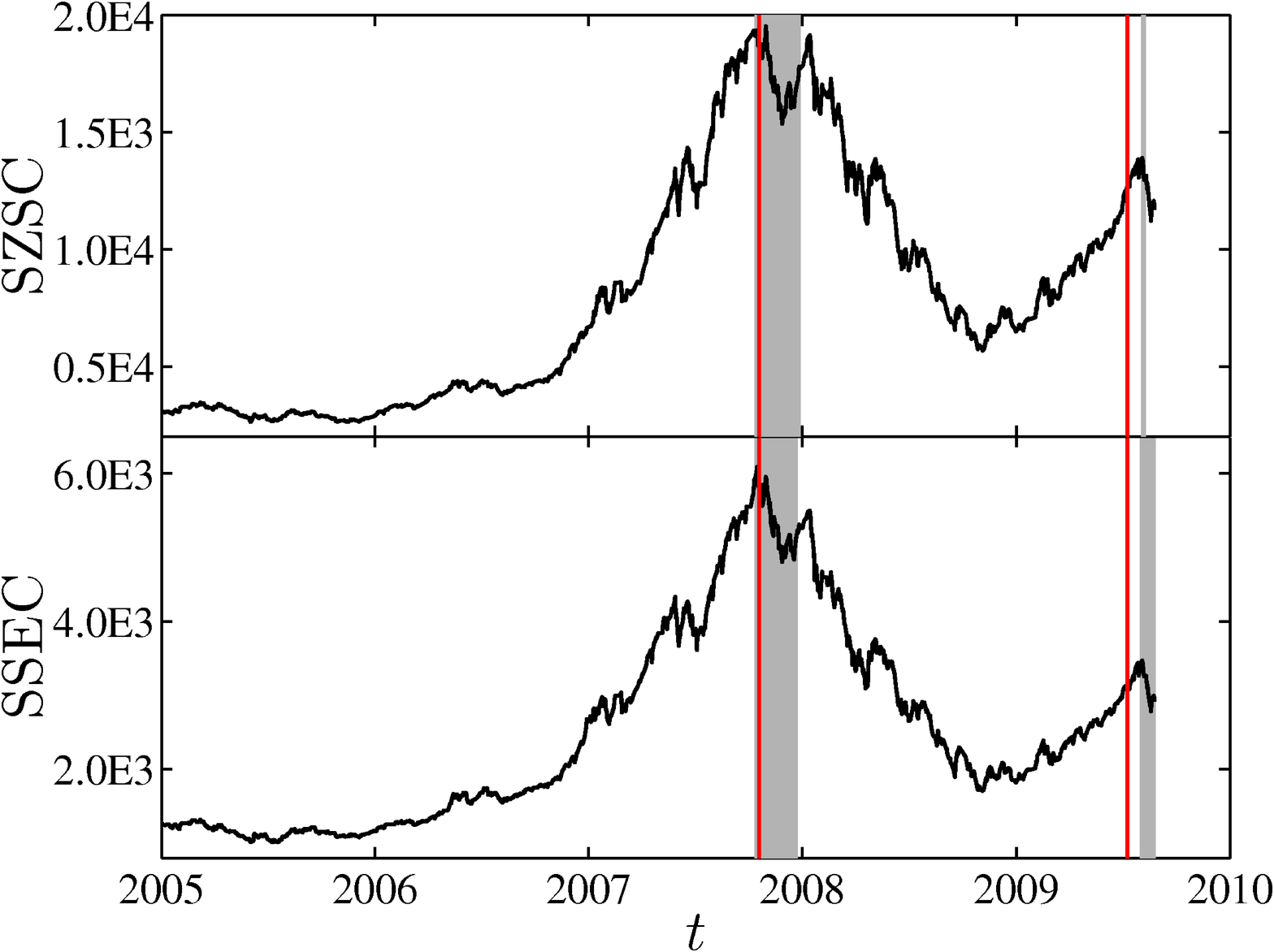}
\caption{\label{Fig:SSEC:SZSC} Evolution of the price trajectories of the SSEC
  index and the SZSC index over the time interval of this analysis. The solid
  red lines indicate the dates of the respective public announcement of our predictions for the two
  bubbles (October 18, 2007 and July 10, 2009) while the grey zones indicate
  the 20\%/80\% confidence intervals for which we forecasted the change of
  regime.  Final closing prices shown in these plots are 10,614.3 (SZSC) and
  2683.72 (SSEC) from September 1, 2009.}
\end{figure}

Here, we present an ex-post analysis of what we identified earlier in their
respective epochs as being two significant bubbles developing in the major
Chinese stock markets, the first one from 2005 to 2007 and the second one from
2008 to 2009.  The organized stock market in mainland China is composed of two
stock exchanges, the Shanghai Stock Exchange (SHSE) and the Shenzhen Stock
Exchange (SZSE). The most important indices for A-shares in SHSE and SZSE are
the Shanghai Stock Exchange Composite index (SSEC) and the Shenzhen Stock
Exchange Component index (SZSC).  The SSEC and SZSC indexes have suffered a
more than $70\%$ drop from their historical high during the period from October
2007 to October 2008. Since November 2008 and until the end of July 2009, the
Chinese stock markets had been rising dramatically.  By calibrating the recent
market index price time series to our LPPL model, we infer that, in both cases,
a bubble had formed in the Chinese stock market and that the market prices were
in an unsustainable state. We present the analysis that led us to diagnose the
presence of these two bubbles respectively in September 2007 and in July 10,
2009 \citep{Bastiaensen-Cauwels-Sornette-Woodard-Zhou-2009-XXX}, and to issue an advance notice of the probable
time of the regime shifts, from a bubble (accelerating ``bullish'') phase to a
(``bearish'') regime or a crash.  See Fig.~\ref{Fig:SSEC:SZSC} for an overview
of the two bubbles and our predictions.  The figure shows the time evolution of
two Chinese indexes, the dates when we made our predictions and the time
intervals of our predicted changes of regime.

The organization of the paper is the following. In Sec.~\ref{sec:methods}, we
present technical descriptions of all the methods used in this paper.
Specifically, they are LPPL fitting procedure, Lomb spectral analysis, unit
root tests and change-of-regime statistic.  We present the results of the 2007
and 2009 bubbles in two separate subsections of Sec.~\ref{sec:results-all}.  In
Sec.~\ref{sec:pred-both-crash}, we document and discuss the predictions we made
for both bubbles prior to their bursting and, further, describe the
observations of both markets indicating that these two bubbles actually did
burst.  Section 5 concludes.

\section{Methods}
\label{sec:methods}

Our main method for detecting bubbles and predicting the critical time $t_c$
when the bubble will end either in a crash or change of regime is by fitting
observed price time series to a log periodic power law (LPPL) model
\citep{Sornette-2003-PR,Sornette-2003,Zhou-2007}. This is a stochastic fitting
procedure that we complement with other techniques, described below.  This
philosophy of using multiple measures aids in filtering predictions, in that a
candidate prediction must pass all tests to be considered worthy.  These
techniques form a toolset that has successfully been put to practice over the
past years by Sornette et al. as described in the introduction. Independently a
similar toolset has recently been developed within the Research Group of BNP
Paribas Fortis (Global Markets) on the same methodology but with a slightly
different implementation of the fitting procedure and the Lomb analysis.

\subsection{General LPPL fitting technique}
\label{sec:general-lppl-fitting}

Consider a time series (such as share price) $p(t)$ between starting and ending
dates $t_1$ and $t_2$.  The LPPL model that we use is
\begin{equation}
  \ln[p(t)] = A + B x^m +C x^m \cos(\omega \ln x + \phi),
 \label{Eq:Landau1}
\end{equation}
where $x = t_c - t$ measures the time to the critical time $t_c$.  For $0<m<1$
and $B<0$ (or $m\leq 0$ and $B>0$), the power law term $Bx^m$ describes the
faster-than-exponential acceleration of prices due to positive feedback
mechanisms. The term proportional to $\cos(\omega\ln x+\phi)$ expresses a
correction to this super-exponential behavior, which has the symmetry of
discrete scale invariance \citep{Sornette-1998-PR}.  By varying $t_1$ and $t_2$,
we can investigate the stability of the fitting parameters with respect to
starting and ending points.

It is worthwhile pointing out that calibrating Eq.~\eqref{Eq:Landau1} to any
given price (or log-price) trajectory will always provide some fit parameters.
That is, any model can be fit to any data.  Hence, it is necessary to establish
a constraint---the LPPL condition---to filter all of the fitting results.  We
filter on three parameters:
\begin{equation}
  \label{eq:LPPL-condition}
  t_c > t_2, B < 0, 0 < m < 1.
\end{equation}
This filter selects regimes with faster-than-exponential acceleration of the
log-price with a diverging slope at the critical future time $t_c$.

There are four nonlinear parameters ($t_c$, $m$, $\omega$, and $\phi$) and
three linear parameters ($A$, $B$, and $C$) in Eq.~\eqref{Eq:Landau1}. In order
to reduce the fitting parameters, the linear parameters are slaved to the
nonlinear parameters. By rewriting Eq.~\eqref{Eq:Landau1} as $\ln
p(t)=A+Bf(t)+Cg(t)$ and using an estimate of the nonlinear parameters, the
linear parameters can be solved analytically via:
\begin{equation}
  \begin{bmatrix}
    N & \sum f_i & \sum g_i \\
    \sum f_i & \sum f_i^2 & \sum g_if_i \\
    \sum g_i & \sum f_ig_i & \sum g_i^2 \\
  \end{bmatrix}
  \begin{bmatrix}
    A\\
    B\\
    C\\
  \end{bmatrix}=
  \begin{bmatrix}
    \sum \ln p_i \\
    \sum \ln p_i f_i \\
    \sum \ln p_i g_i \\
  \end{bmatrix}.
  \label{Eq:LPPL:Fit}
\end{equation}
The implementation of the fitting proceeds in two steps. First, we adopt the
Taboo search \citep{Cvijovic-Klinowski-1995-Science} to find 10 candidate solutions
from our given search space. Second, each of these solutions is used as an
initial estimate in a Levenberg-Marquardt nonlinear least squares
algorithm. The solution with the minimum sum of squares between model and
observations is taken as the solution.

\subsection{Stability of fits vs. shrinking and expanding intervals
and probabilistic forecasts}
\label{sec:stability-fits-vs}

In order to test the sensitivity of variable fitting intervals $[t_1, t_2]$, we
adopt the strategy of fixing one endpoint and varying the other one. For
instance, if $t_2$ is fixed, the time window shrinks in terms of $t_1$ moving
towards $t_2$ with a step of five days. If $t_1$ is fixed, the time window
expands in terms of $t_2$ moving away from $t_1$ with a step of five days. For
each such $[t_1, t_2]$ interval, the fitting procedure is implemented on the
index series three times.  Recall that because of the rough nonlinear parameter
landscape of Eq.~\eqref{Eq:Landau1} and the stochastic nature of our initial
parameter selection, it is expected that each implementation of our fit process
will produce a different set of fit parameters. By repeating the process
multiple times, we investigate an optimal (not necessarily the
optimal) region of solution space.

By sampling many intervals as well as by using bootstrap techniques, we obtain
predictions that are inherently probabilistic and reflect the intrinsic noisy
nature of the underlying generating processes.  This allows us to provide
probabilistic estimations on the time intervals in which a given bubble may end
and lead to a new market regime.  In this respect, we stress that,
notwithstanding the common use of the term ``crash'' to refer to the aftermath
of a bubble, a real crash does not always occurs. Rather, the end of a bubble
may be the most probable time for a crash to occur, but the bubble may end
without a splash and, instead, transition to a plateau or a slower decay. This
point is actually crucial in rational expectation models of bubbles in that,
even in the presence of investors fully informed of the presence of the bubble
and with the knowledge of its end date, it remains rational to stay invested in
the market to garner very large returns since the risk of a crash remains
finite \citep{Johansen-Sornette-Ledoit-1999-JR,Johansen-Ledoit-Sornette-2000-IJTAF}.

\subsection{Lomb spectral analysis}
\label{sec:lomb-spectr-analys}

Fitting the logarithm of prices to the model Eq.~\eqref{Eq:Landau1} gives
strong evidence supporting log-periodicity in that stable values of the angular
frequency $\omega$ are found.  We test this feature further by using Lomb
spectral analysis \citep{Press-Teukolsky-Vetterling-Flannery-1996} for detecting
the log-periodic oscillations.  The Lomb method is a spectral analysis designed
for irregularly sampled data and gives the same results as the standard Fourier
spectral analysis for evenly spaced data.  Specifically, given a time series,
the Lomb analysis returns a series of frequencies $\omega$ and associated power
at each frequency, $P_N(\omega))$.  The frequency with the maximum power is
taken as the Lomb frequency, $\omega_{\rm{Lomb}}$.  Following
\citet{Sornette-Zhou-2002-QF}, the spectral Lomb analysis is performed on two
types of signals.

\paragraph{Parametric detrending approach}
\label{sec:param-detr-appr}

The first is the series of detrended residuals, calculated as
\begin{equation}
  r(t) = x^{-m}(\ln[p(t)] - A - B x^m),
  \label{Eq:residual}
\end{equation}
where $x \equiv t_c -t$ and $A$, $B$, $t_c$ and $m$ have been found via the
method of Section~\ref{sec:general-lppl-fitting}
\citep{Johansen-Sornette-1999-Risk,Zhou-Jiang-Sornette-2007-PA}.  As suggested
in Eq.~\eqref{Eq:Landau1}, the log-periodic oscillations result from the cosine
part.  The angular frequency $\omega_{\rm{Lomb}}$ is then compared with that
found in the LPPL fitting procedure, $\omega_{\rm{fit}}$.

\paragraph{Non-parametric, (H, q) analysis}
\label{sec:non-parametric-h}

The second is the $(H,q)$-derivative of the logarithmic prices, which has been
successfully applied to financial crashes \citep{Zhou-Sornette-2003-IJMPC} and
critical ruptures \citep{Zhou-Sornette-2002-PRE} for the detection of
log-periodic components.  The $(H,q)$ analysis is a generalization of the
$q$-analysis \citep{Erzan-1997-PLA,Erzan-Eckmann-1997-PRL}, which is a natural
tool for the description of discrete scale invariance. The $(H,q)$-derivative
is defined as,
\begin{equation}
  D_q^H f(x) \triangleq \frac{f(x)-f(qx)}{[(1-q)x]^H}.
  \label{Eq:Hq}
\end{equation}
We vary $H$ and $q$ in the ranges [-1, 1] and [0, 1], respectively, and perform
the Lomb analysis on the resulting series.  If $H=1$ in Eq.~\eqref{Eq:Hq}, the
$(H,q)$-derivative reduces to the normal $q$-derivative, which itself reduces
to the normal derivative in the limit $q \rightarrow 1^-$. Without loss of
generality, $q$ is constrained in the open interval $(0,1)$. The advantage of
the $(H,q)$ analysis is that there is no need for detrending, as it is
automatically accounted for by the finite difference and the normalization by
the denominator. This method has been applied for detecting log-periodicity in
stock market bubbles and anti-bubbles
\citep{Sornette-Zhou-2002-QF,Zhou-Sornette-2003-IJMPC,Zhou-Sornette-2004a-PA},
in the USA foreign capital inflow bubble ending in early 2001
\citep{Sornette-Zhou-2004-PA} and in the ongoing UK real estate bubble
\citep{Zhou-Sornette-2003a-PA}.

\subsection{Ornstein-Uhlenbeck and unit root tests}
\label{sec:ornstein-uhlenbeck}

Recently, \citet{Lin-Ren-Sornette-2009-XXX} have put forward a self-consistent model for explosive
financial bubbles, in which the LPPL fitting
residuals can be modeled by a mean-reversal Ornstein-Uhlenbeck (O-U) process if
the logarithmic price in the bubble regime is attributed to a deterministic
LPPL component. The test for the O-U property of LPPL fitting residuals can be
translated into an AR(1) test for the corresponding residuals. Hence, we can
verify the O-U property of fitting residuals by applying unit-root tests on the
residuals.  We use the Phillips-Perron and Dickey-Fuller unit-root tests. A
rejection of null hypothesis $H_0$ suggests that the residuals are stationary
and therefore are compatible with the O-U process in the residuals. Our tests
use the same time windows as the LPPL calibrating procedure.

\subsection{Statistic of change of regime \label{y3ykbwgw}}

We define a simple statistic to demonstrate the change of regime in the
``post-mortem'' analysis of our prediction.  We calculate the difference
$d_{co}(t)$ between the closing and opening prices on each trading day and
track the fraction of days with negative $d_{co}(t)$ in a rolling window of T
days, with $T=10, 20$ and $30$ days.

\section{Results}
\label{sec:results-all}

In the following two subsections, we present our analysis of the separate 2007
and 2009 bubbles using the methods described in Section~\ref{sec:methods}.

\subsection{Back test of Chinese bubble from 2005 to 2007}
\label{sec:results-bubble-2007}

\subsubsection{LPPL fitting with varying window sizes}
\label{sec:results-lppl-fits}

\begin{figure}[htp]
\centering
\includegraphics[width=7cm]{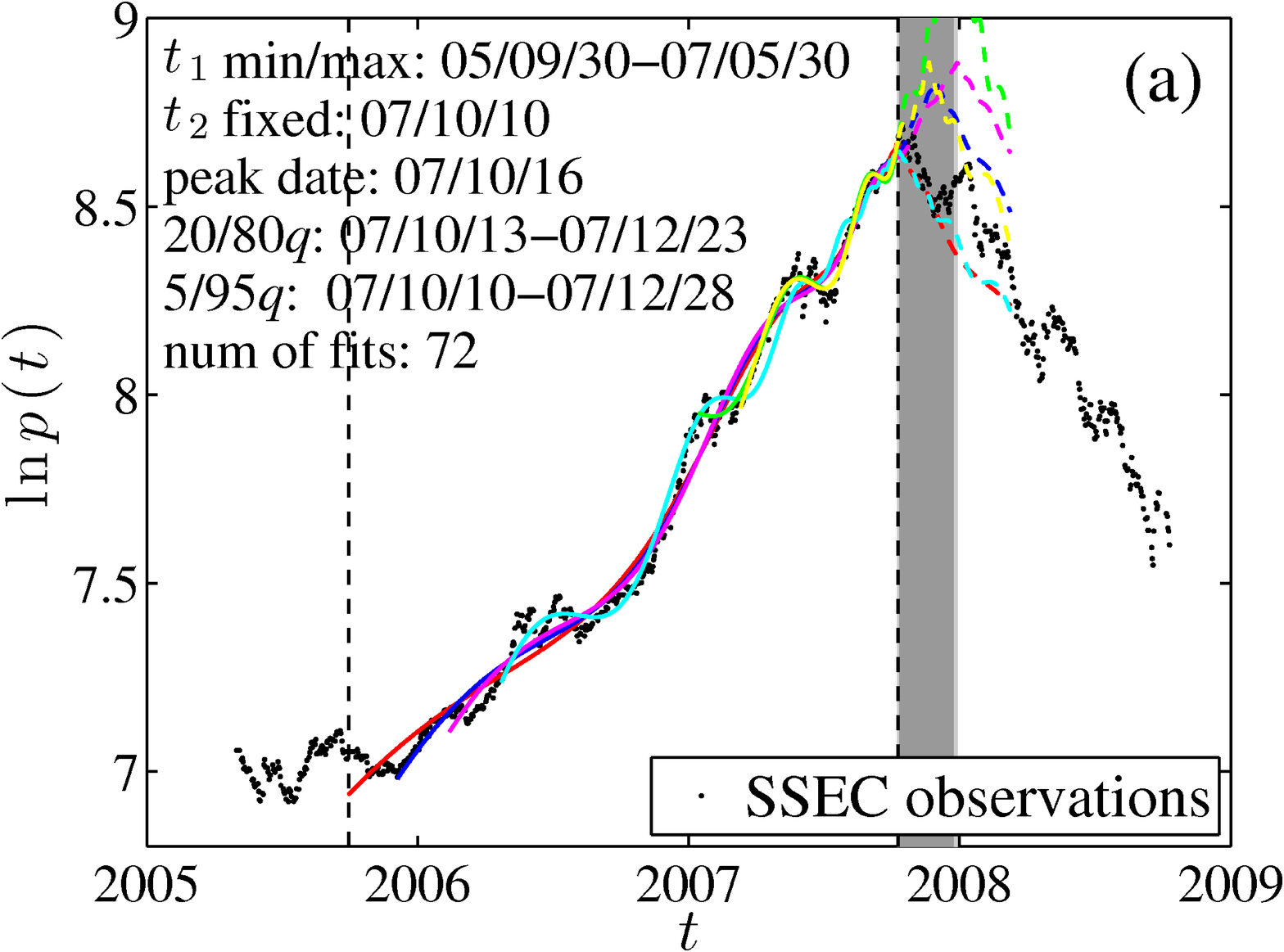}
\includegraphics[width=7cm]{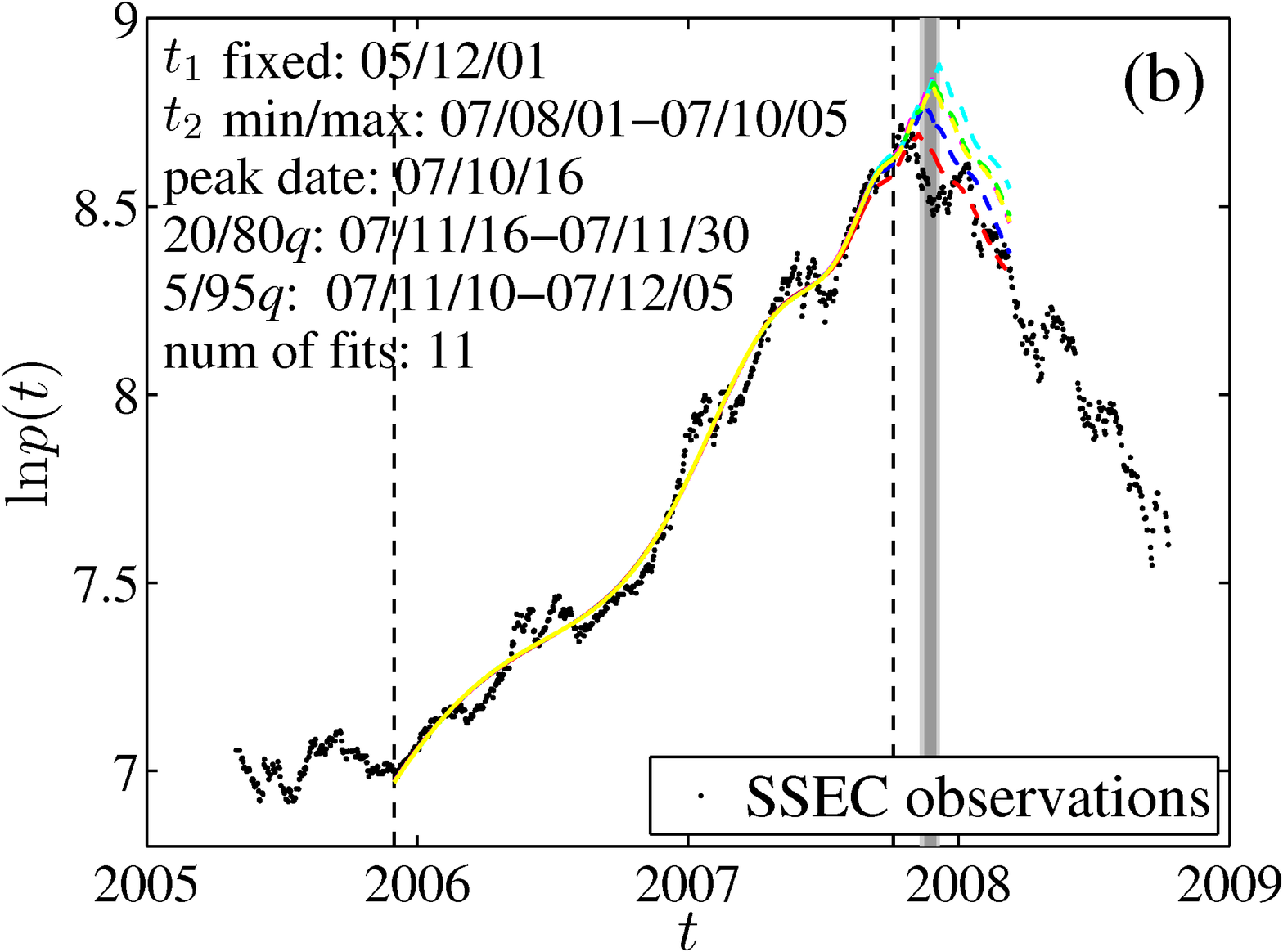}
\includegraphics[width=7cm]{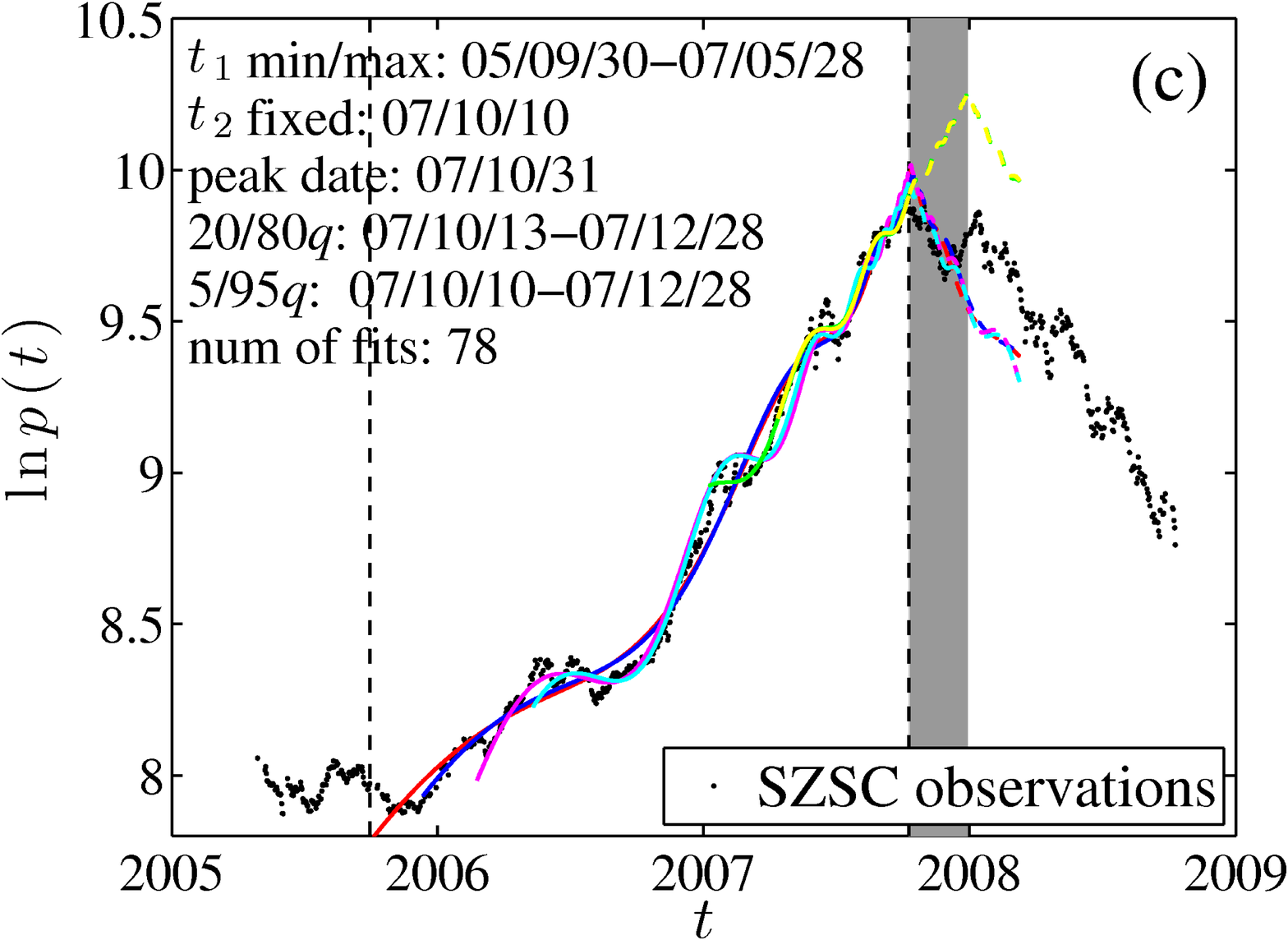}
\includegraphics[width=7cm]{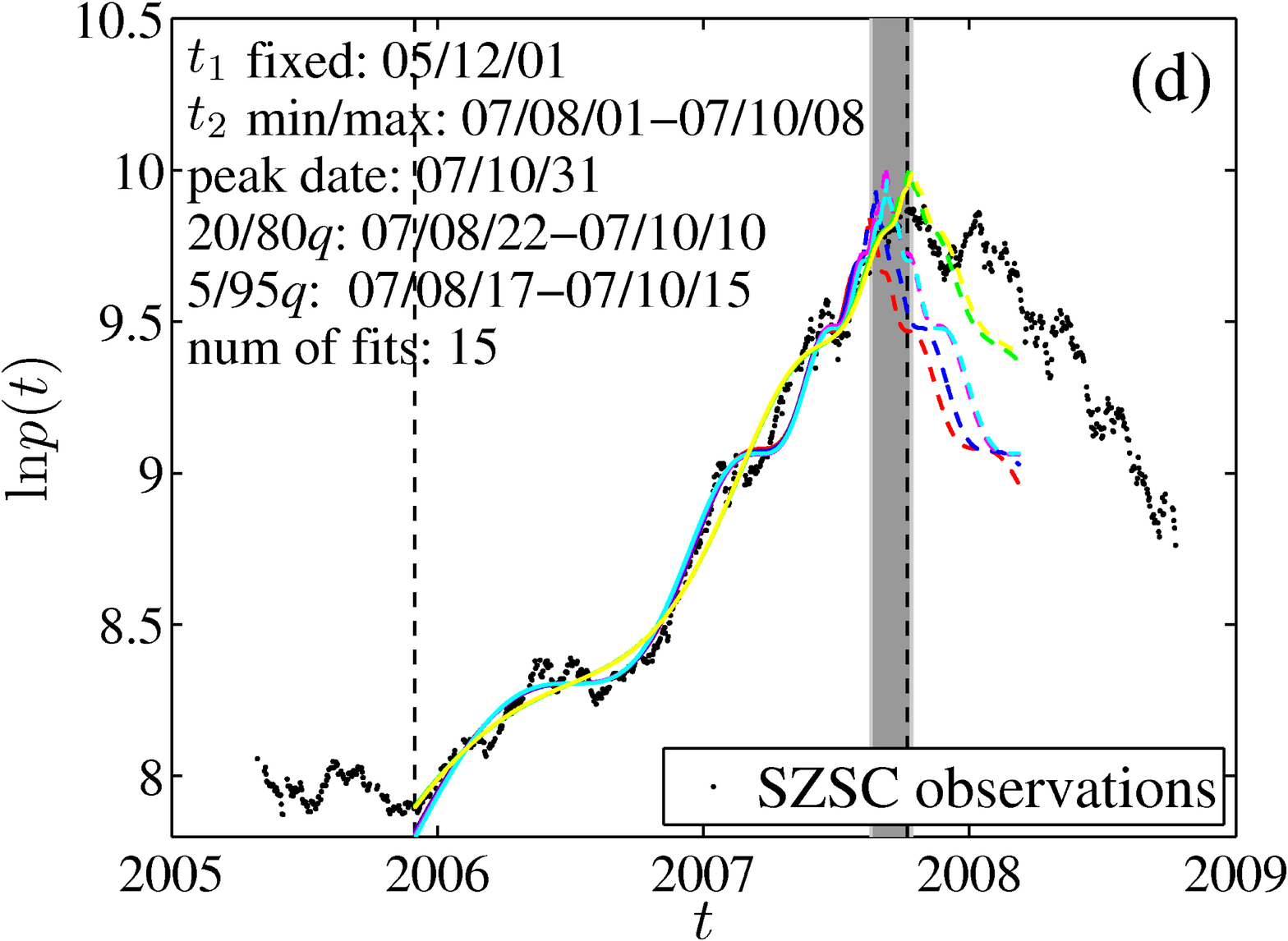}
\caption{\label{Fig:B0507} Daily trajectory of the logarithmic SSEC (a,b) and
  SZSC (c,d) index from May-01-2005 to Oct-18-2008 (dots) and fits to the LPPL
  formula \eqref{Eq:Landau1}. The dark and light shadow box indicate 20/80\%
  and 5/95\% quantile range of values of the crash dates for the fits,
  respectively. The two dashed lines correspond to the minimum date of $t_1$
  and the maximum date of $t_2$. (a) Examples of fitting to shrinking windows
  with varied $t_1$ and fixed $t_2$ = Oct-10-2007 for SSEC. The six fitting
  illustrations are corresponding to $t_1$ = Sep-30-2005, Dec-05-2005,
  Feb-13-2006, Apr-24-2006, Jan-15-2007, and Mar-12-2007. (b) Examples of
  fitting to expanding windows with fixed $t_1$ = Dec-01-2005 and varied $t_2$
  for SSEC. The six fitting illustrations are associated with $t_2$ =
  Aug-20-2007, Aug-29-2007, Sep-07-2007, Sep-17-2007, Sep-26-2007,
  Oct-05-2007. (c) Examples of fitting to shrinking windows with varied $t_1$
  and fixed $t_2$ = Oct-10-2007 for SZSC.  The six fitting illustrations are
  corresponding to $t_1$ = Sep-30-2005, Dec-12-2006, Feb-24-2006, May-12-2006,
  Jan-09-2007, and Apr-13-2007. (d) Examples of fitting to expanding windows
  with fixed $t_1$ = Dec-01-2005 and varied $t_2$ for SZSC. The six fitting
  illustrations are associated with $t_2$ = Aug-01-2007, Aug-10-2007,
  Aug-24-2007, Sep-07-2007, Sep-21-2007, Oct-08-2007.}
\end{figure}

As discussed above, we test the stability of fit parameters for the two
indexes, SSEC and SZSC, by varying the size of the fit intervals.
Specifically, the logarithmic index is fit by the LPPL formula,
Eq.~\eqref{Eq:Landau1}:
\begin{enumerate}
\item in shrinking windows with a fixed end date $t_2 =$ Oct-10-2007 with the
  start time $t_1$ increasing from Oct-01-2005 to May-31-2007 in steps of five
  (trading) days and
\item in expanding windows with a fixed start date $t_1$ = Dec-01-2005 with the
  end date $t_2$ increasing from May-01-2007 to Oct-01-2007 in steps of five
  (trading) days.
\end{enumerate}
In the above two fitting procedures, we fit the indexes 124 times in shrinking
windows and 15 times in expanding windows. After filtering by the LPPL
conditions, we finally observe 72 (78) results in the first step and 11 (15)
results in the second step for SSEC (respectively
SZSC). Figures~\ref{Fig:B0507} (a) and (c) illustrate six chosen fitting
results of the shrinking windows for SSEC and SZSC and Figures~\ref{Fig:B0507}
(b) and (d) illustrate six fitting results of the expanding time intervals for
SSEC and SZSC. The dark and light shadow boxes in the figures indicate
20\%/80\% and 5\%/95\% quantile range of values of the crash dates for the fits
that survived filtering. One can observe that the observed market peak dates
(16~October~2007 for SSEC, 31~October~2007 for SZSC) lie in the quantile ranges
of predicted crash dates $t_c$ using only data from before the market crash
(i.e., using $t_2 < t_{c\_\rm{obs}}$).

\subsubsection{Lomb analysis, parametric approach}
\label{sec:results-lomb}

\begin{figure}[htp]
\centering
\includegraphics[width=7cm]{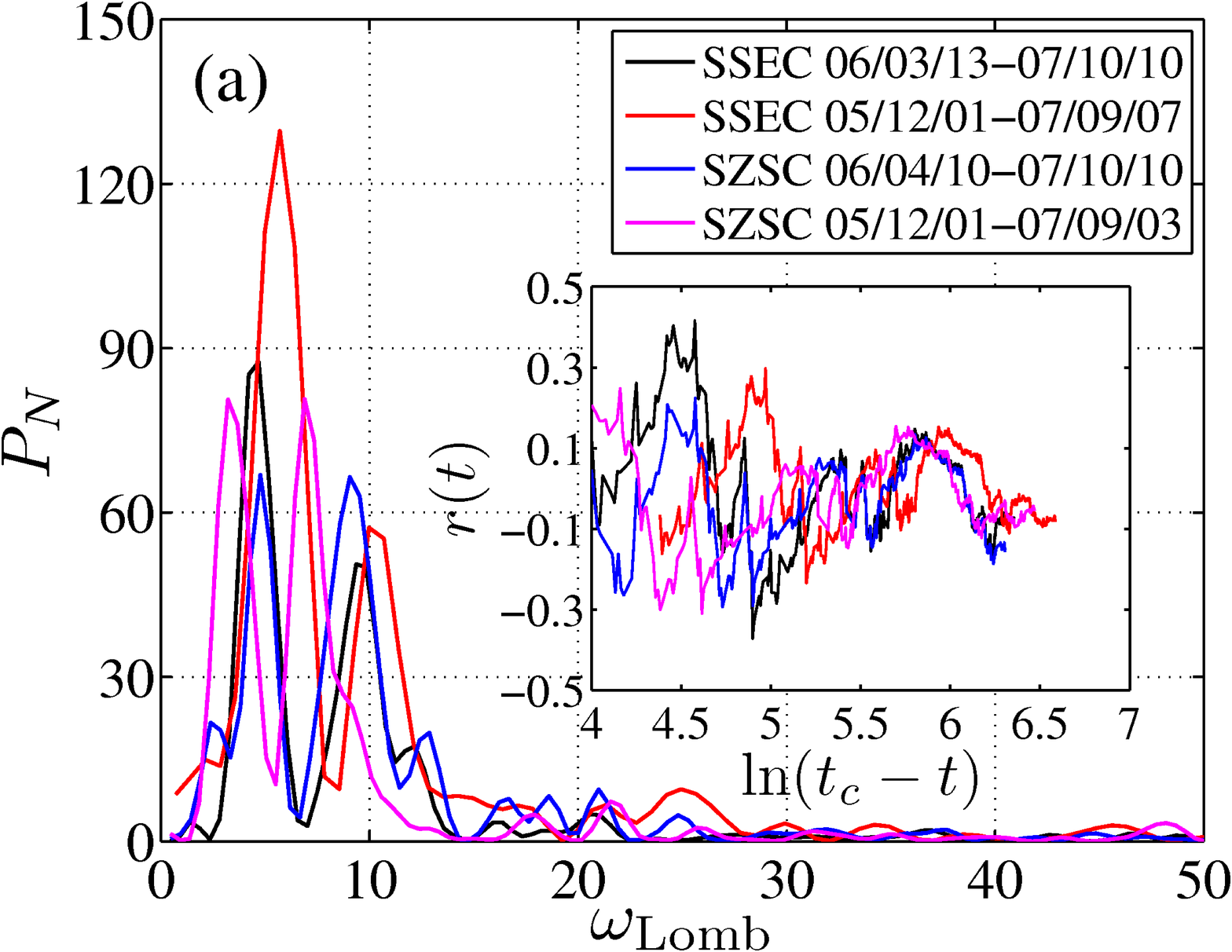}
\includegraphics[width=7cm]{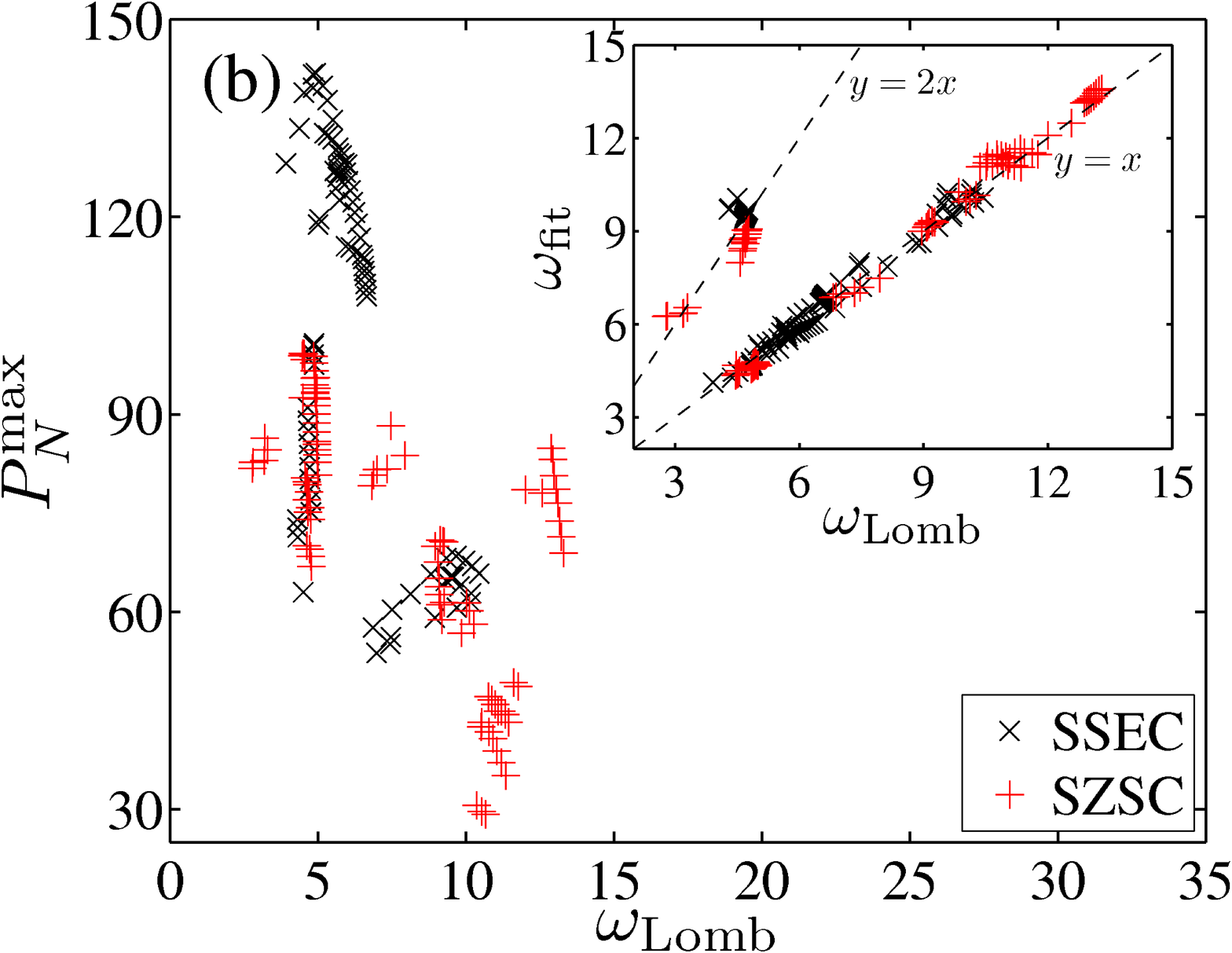}
\caption{\label{Fig:Lomb:r} Lomb tests of the detrending residuals $r(t)$ for
  SSEC and SZSC. The residuals are obtained from Eq.~(\ref{Eq:residual}) by
  substituting different survival LPPL calibrating windows with the
  corresponding fitting results including $t_c$, $m$, and $A$. (a) Lomb
  periodograms for four typical examples, which are presented in the
  legend. The time periods followed the index names represent the LPPL
  calibrating windows. The inset illustrates the corresponding residuals $r(t)$
  as a function of $\ln(t_c-t)$. (b) Bivariate distribution of pairs
  $(\omega_{\rm{Lomb}}, P_N^{\max})$ for different LPPL calibrating
  intervals. Each point in the figure stands for the highest peak and its
  associated angular log-frequency in the Lomb periodogram of a given detrended
  residual series. The inset shows $\omega_{\rm{fit}}$ as a function of
  $\omega_{\rm{Lomb}}$.}
\end{figure}

Fig.~\ref{Fig:Lomb:r} summarizes the results of our Lomb analysis on the
detrended residuals $r(t)$. The Lomb periodograms ($P_N$ with respect to
$\omega_{\rm{Lomb}}$) are plotted in Fig.~\ref{Fig:Lomb:r}(a) for four typical
examples, which are ($t_1, t_2$) = (Mar-13-2006, Oct-10-2007) and (Dec-12-2005,
Sep-07-2007) for SSEC and (Apr-04-2006, Oct-10-2007) and (Dec-01-2005,
Sep-09-2007) for SZSC. The inset illustrates the corresponding detrended
residuals $r(t)$ as a function of $\ln(t_c-t)$. We select the highest peak with
its associated $\omega_{\rm{Lomb}}$.

The values of $\omega_{\rm{Lomb}}$ must be consistent with the values of
$\omega_{\rm{fit}}$ obtained from the fitting.  We plot the bivariate
distribution of pairs $(\omega_{\rm{Lomb}}, P_N^{\max})$ for different LPPL
calibrating windows in Fig.~\ref{Fig:Lomb:r}(b) and find that the minimum value
of $P_N^{\rm{\max}}$ is approximately 54 for SSEC and approximately 30 for
SZSC. These peaks are linked to a false alarm probability, which is defined as
the chance that we falsely detect log-periodicity in a signal without true
log-periodicity.

To calculate this false alarm probability, a model of the distribution of the
residuals must be used.  We `bracket' a range of models, from uncorrelated
white noise to long-range correlated noise.  For white noise, we find the false
alarm probability to be $Pr \ll 10^{-5}$
\citep{Press-Teukolsky-Vetterling-Flannery-1996}. If the residuals have
power-law behaviors with exponent in the range 2-4 and long-range correlations
characterized by a Hurst index $H \leq 0.7$, we have $Pr < 10^{-2}$
\citep{Zhou-Sornette-2002-IJMPC}.

The inset of Fig.~\ref{Fig:Lomb:r}(b) plots $\omega_{\rm{fit}}$ with respect to
$\omega_{\rm{Lomb}}$. One can see that most pairs of $(\omega_{\rm{Lomb}},
\omega_{\rm{fit}})$ are overlapping on the line $y=x$, which indicates the
consistency between $\omega_{\rm{fit}}$ and $\omega_{\rm{Lomb}}$. The other
pairs are located on the line $y=2x$. We can interpret these as a fundamental
log-periodic component at $\omega$ and its harmonic component at $2\omega$.
The existence of harmonics of log-periodic components can be expected
generically in log-periodic signals
\citep{Sornette-1998-PR,Gluzman-Sornette-2002-PRE,Zhou-Sornette-2002-PD,Zhou-Sornette-2009b-PA} and has been
documented in earlier studies both of financial time series and for other
systems \citep{Zhou-Jiang-Sornette-2007-PA}. When the harmonics are well defined
with close-to-integer ratios to a common fundamental frequency as is the case
here, this is in general a diagnostic of a very significant log-periodic
component.

\subsubsection{Lomb analysis, non-parametric $(H, Q)$ approach}
\label{sec:results-hq}

\begin{figure}[htp]
\centering
\includegraphics[width=7cm]{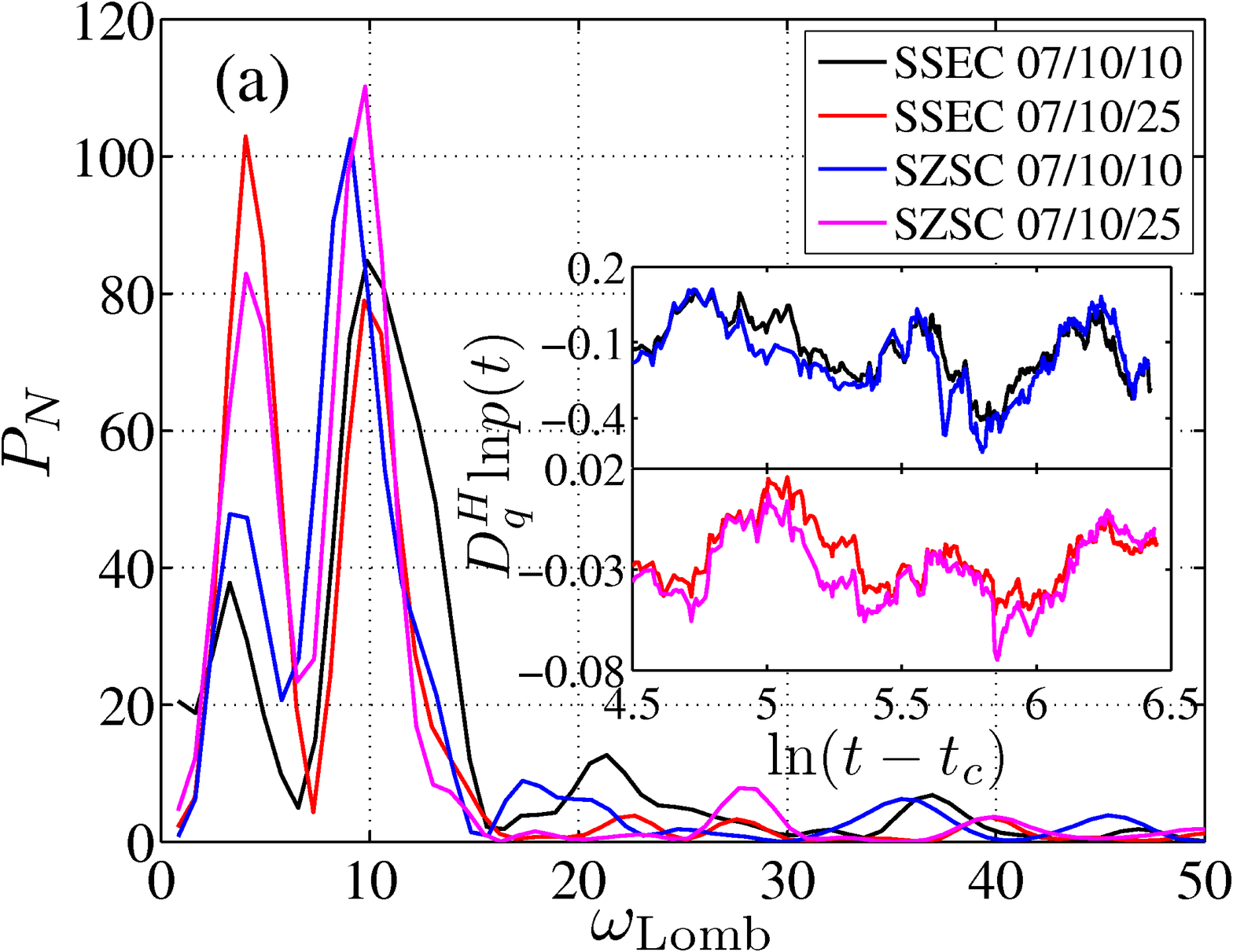}
\includegraphics[width=7cm]{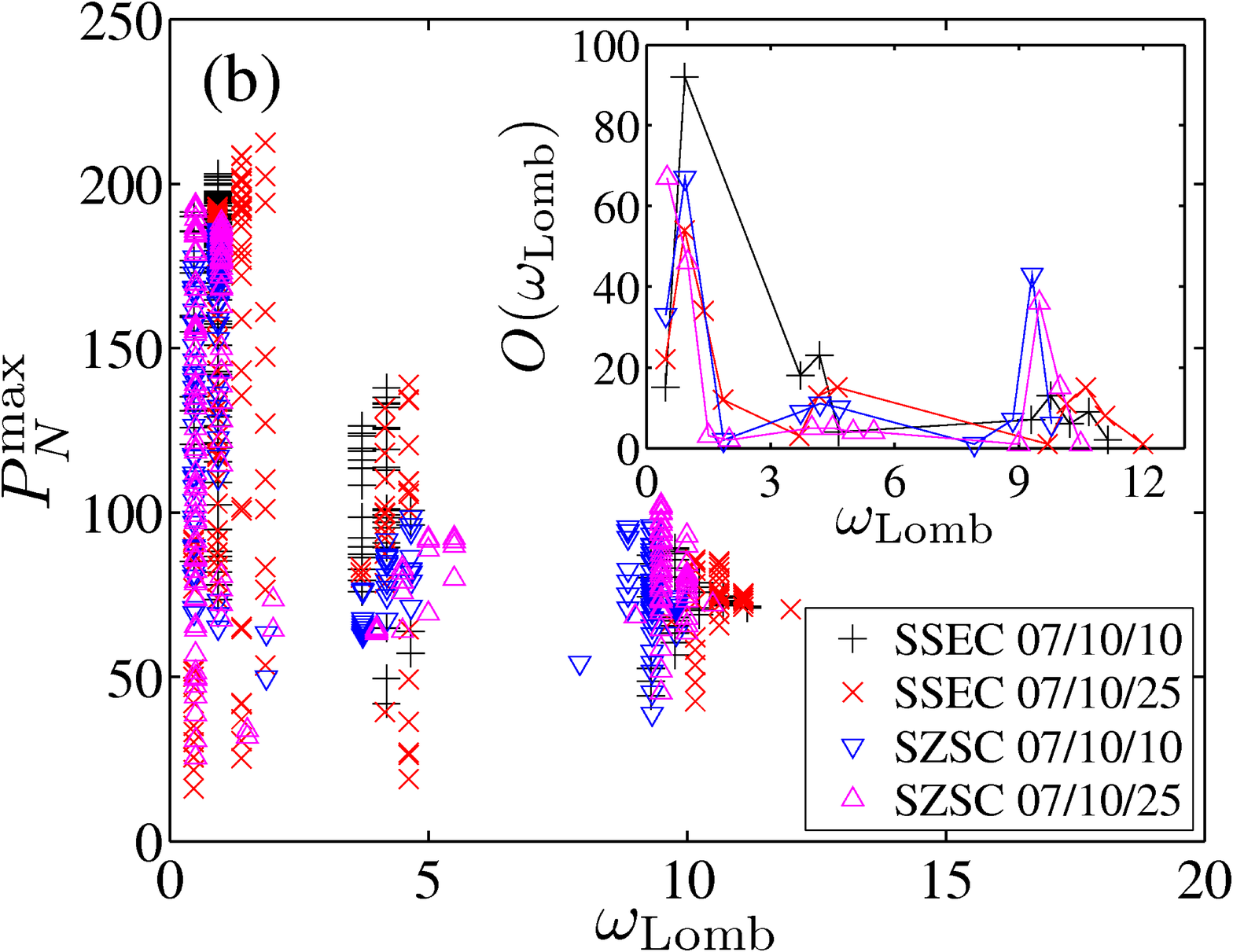}
\caption{\label{Fig:Lomb:Hq} Lomb tests of $(H,q)$-derivative of logarithmic
  indexes. (a) Lomb periodograms of $D_q^H\ln p(t)$ for four typical examples,
  which are $t_c$ = Oct-10-2007 with $H = 0$ and $q = 0.8$ for SSEC, $t_c$ =
  Oct-25-2007 with $H = 0.5$ and $q = 0.7$ for SSEC, $t_c$ = Oct-10-2007 with
  $H = 0$ and $q = 0.8$ for SZSC, and $t_c$ = Oct-25-2007 with $H = 0.5$ and $q
  = 0.7$ for SZSC, respectively. The inset shows the corresponding plots of
  $D_q^H\ln p(t)$ as a function of $\ln(t_c-t)$. (b) Bivariate distribution of
  pairs $(\omega_{\rm{Lomb}},P_N^{\max})$ for different pairs of $(H,q)$. Each
  point corresponds the highest Lomb peak and its associated angular
  log-frequency in the Lomb periodogram of the $(H,q)$-derivative of
  logarithmically indexes for a given pair $(H,q)$. The inset shows the
  empirical frequency distribution of $\omega_{\rm{Lomb}}$.}
\end{figure}

In order to non-parametrically check the existence of log-periodicity by means
of $(H,q)$-analysis, we take $f(x) = \ln p(t)$ and $x = t_c-t$ with $t_c=$
Oct-10-2007 or Oct-25-2007 (the two observed peak dates of the indexes). For
each pair of $(H,q)$ values, we calculate the $(H,q)$-derivative with a given
$t_c$, on which we calculate the Lomb analysis. Fig.~\ref{Fig:Lomb:Hq}(a)
illustrates the Lomb periodograms for both indexes with $t_c=$ Oct-10-2007,
$H=0$, $q=0.8$ and $t_c=$ Oct-25-2007, $H=0.5$, $q=0.7$. The corresponding
$D_q^H \ln p(t)$, defined by formula (\ref{Eq:Hq}), is plotted with respect to
$\ln(t_c-t)$ in the inset. The highest Lomb peak of the resultant periodogram
has height $P_N^{\max}$ and abscissa $\omega_{\rm{Lomb}}$, both $P_N^{\max}$
and $\omega_{\rm{Lomb}}$ being functions of $H$ and $q$.

We scan a 21 $\times$ 9 rectangular grid in the $(H,q)$ plane, with $H \in [-1,
1]$ and $q \in [0.1, 0.9]$, both in steps of 0.1. Fig.~\ref{Fig:Lomb:Hq}(b)
illustrates the bivariate distribution of pairs $(\omega_{\rm{Lomb}},
P_N^{\max})$. The inset shows a simple histogram of $\omega_{\rm{Lomb}}$. We
observe the three most prominent clusters corresponding to SSEC with $t_c=$
Oct-10-2007 (Oct-25-2007) as $\omega_{\rm{Lomb}}^0 = 0.86 \pm 0.16$ ($1.06 \pm
0.41$), $\omega_{\rm{Lomb}}^1 = 4.04 \pm 0.29$ ($4.33 \pm 0.31$) and
$\omega_{\rm{Lomb}}^2 = 10.05 \pm 0.56$ ($10.59 \pm 0.44$).  For SZSC with
$t_c=$ Oct-10-2007 (Oct-25-2007), we find the three most prominent clusters to
be $\omega_{\rm{Lomb}}^0 = 0.80 \pm 0.27$ ($0.75 \pm 0.32$),
$\omega_{\rm{Lomb}}^1 = 4.21 \pm 0.38$ ($4.33 \pm 0.31$), and
$\omega_{\rm{Lomb}}^2 = 9.29 \pm 0.29$ ($9.65 \pm 0.27$).

The small value of $\omega_{\rm{Lomb}}^0$ corresponds to a component with less
than one full period within the interval of the $\ln(t_c -t)$ variable
investigated here.  According to extensive tests performed in synthetic time
series \citep{Huang-Johansen-Lee-Saleur-Sornette-2000-JGR}, it should be interpreted as a spurious peak
associated with the most probable partial oscillations of a noisy signal and/or
to a residual global trend in the $(H,q)$-derivative.  Then,
$\omega_{\rm{Lomb}}^1$ is identified as the fundamental angular log-frequency
and $\omega_{\rm{Lomb}}^2 \approx 2\omega_{\rm{Lomb}}^1$ is interpreted as its
second harmonic.

\subsection{Back test of Chinese bubble from 2008 to 2009}
\label{sec:results-bubble-2009}

\subsubsection{LPPL fitting with varying window sizes}
\label{sec:2009-lppl-fitting-with}

\begin{figure}[htp]
\centering
\includegraphics[width=7cm]{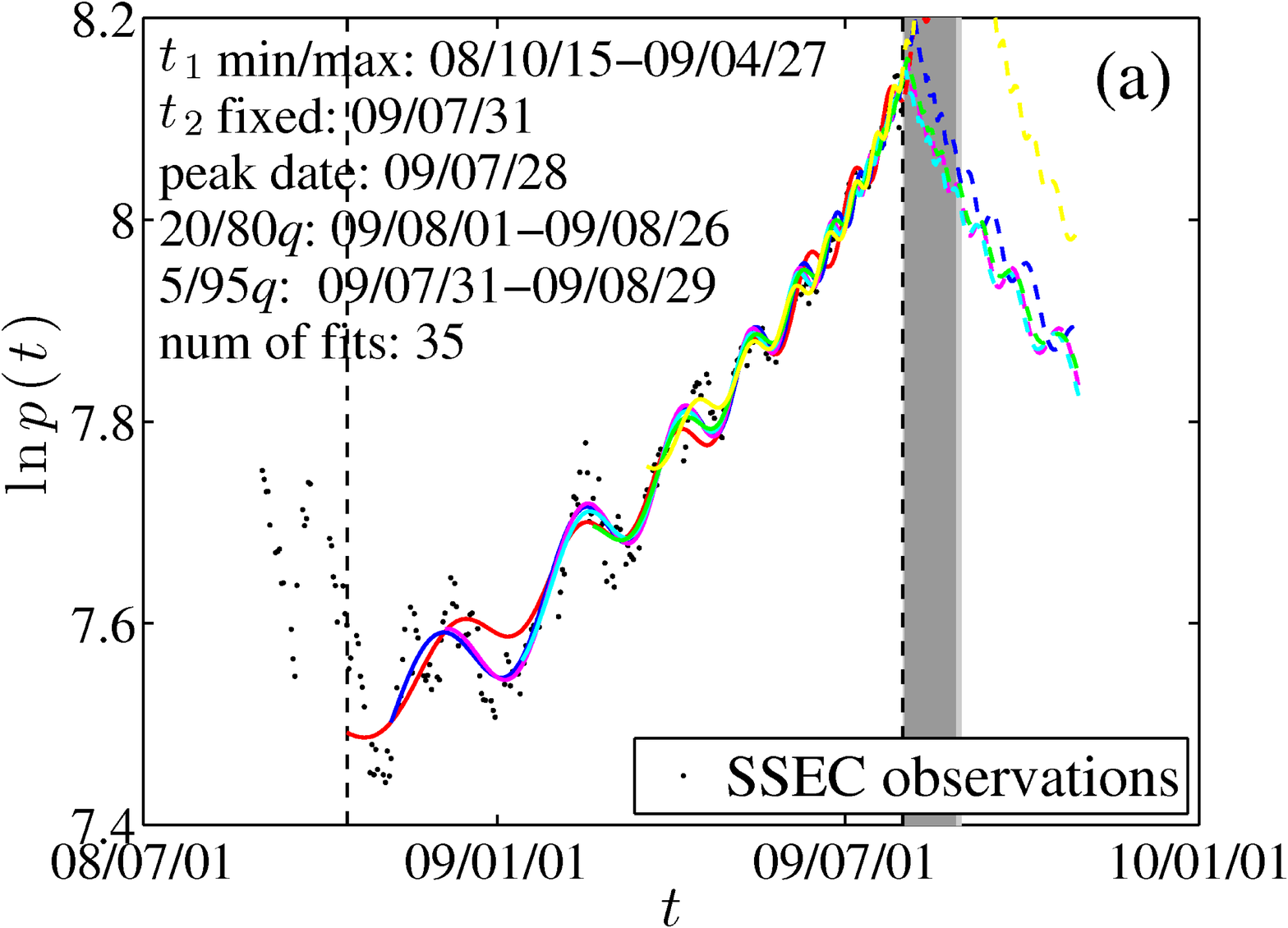}
\includegraphics[width=7cm]{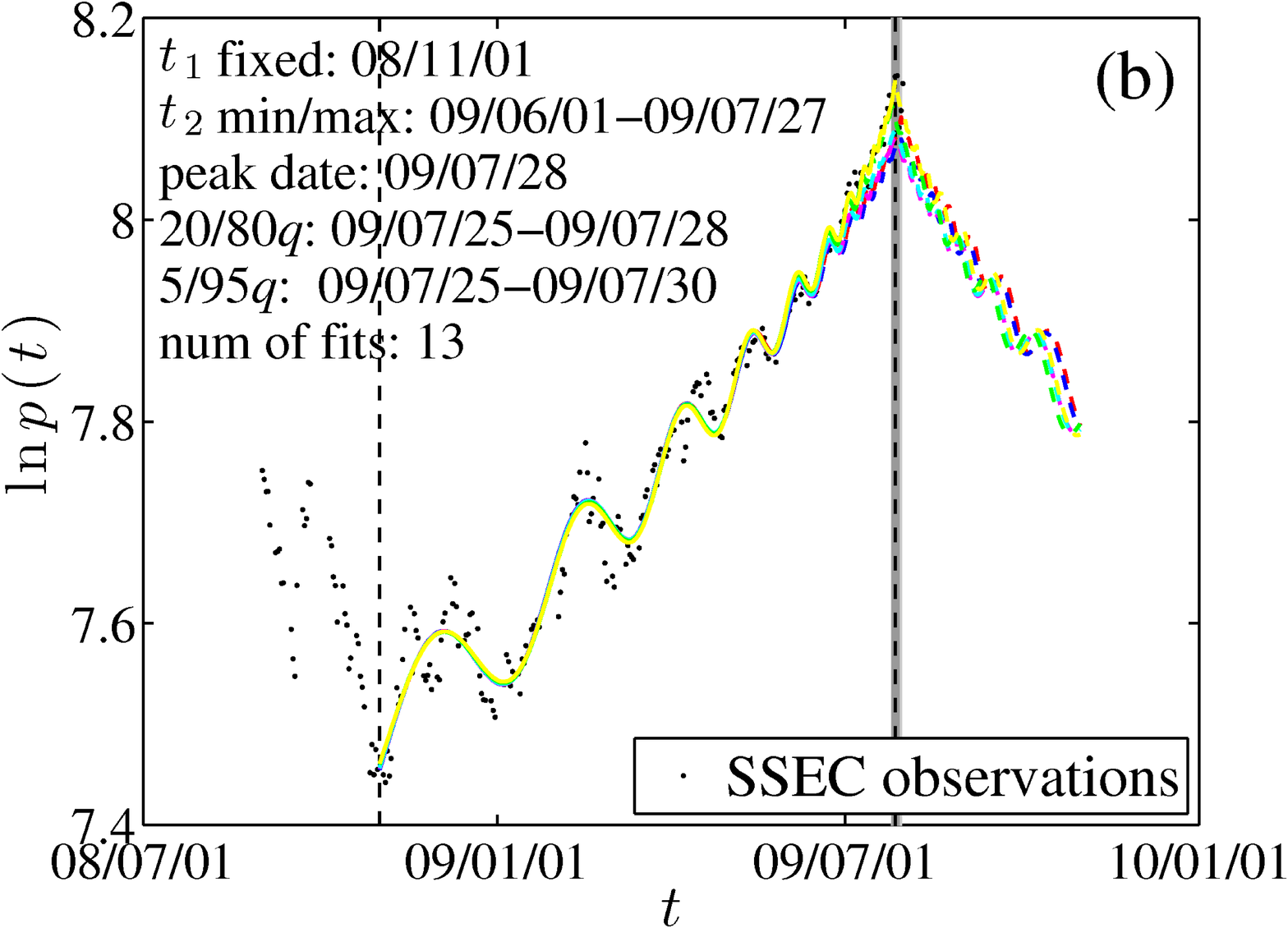}
\includegraphics[width=7cm]{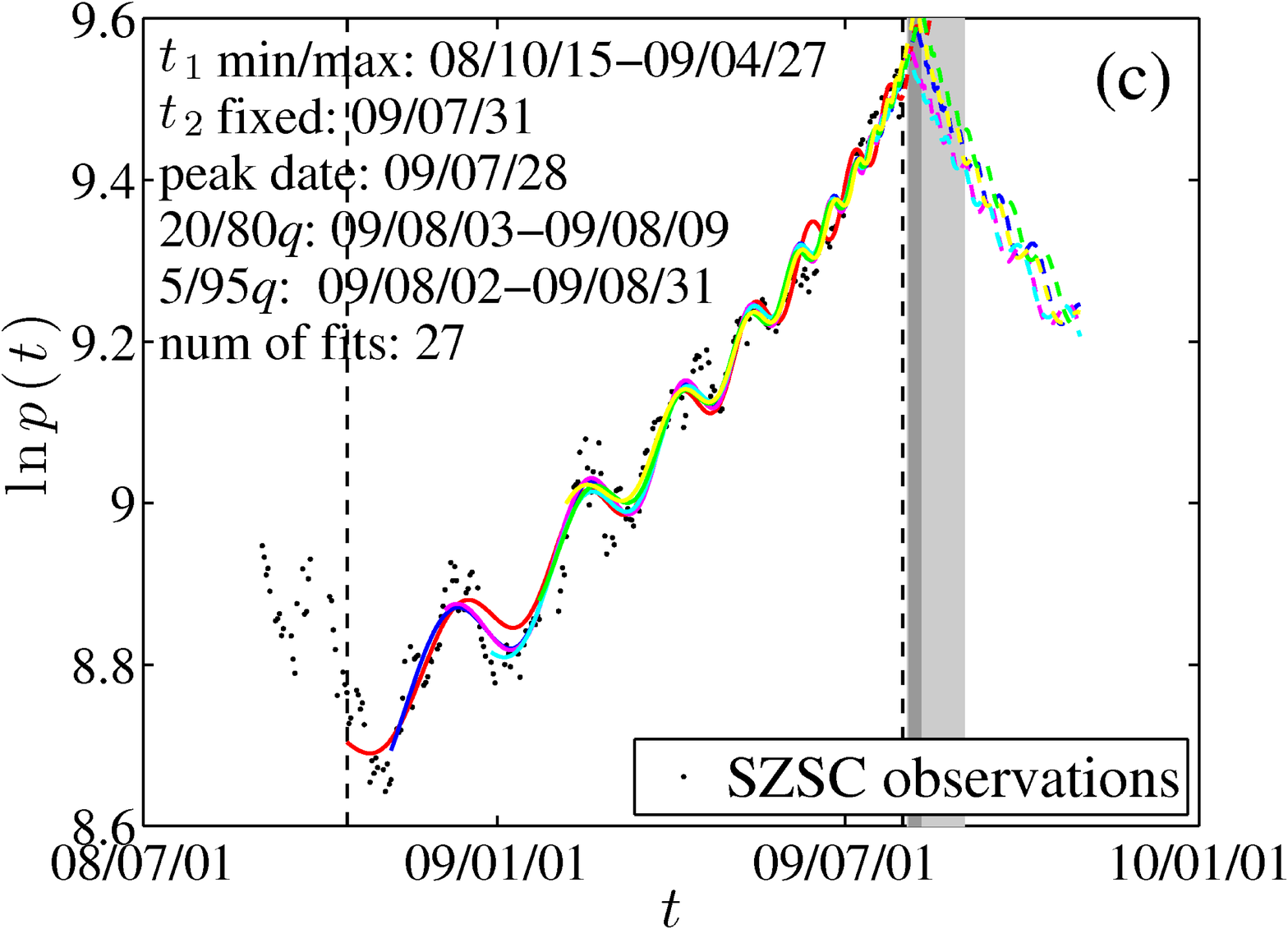}
\includegraphics[width=7cm]{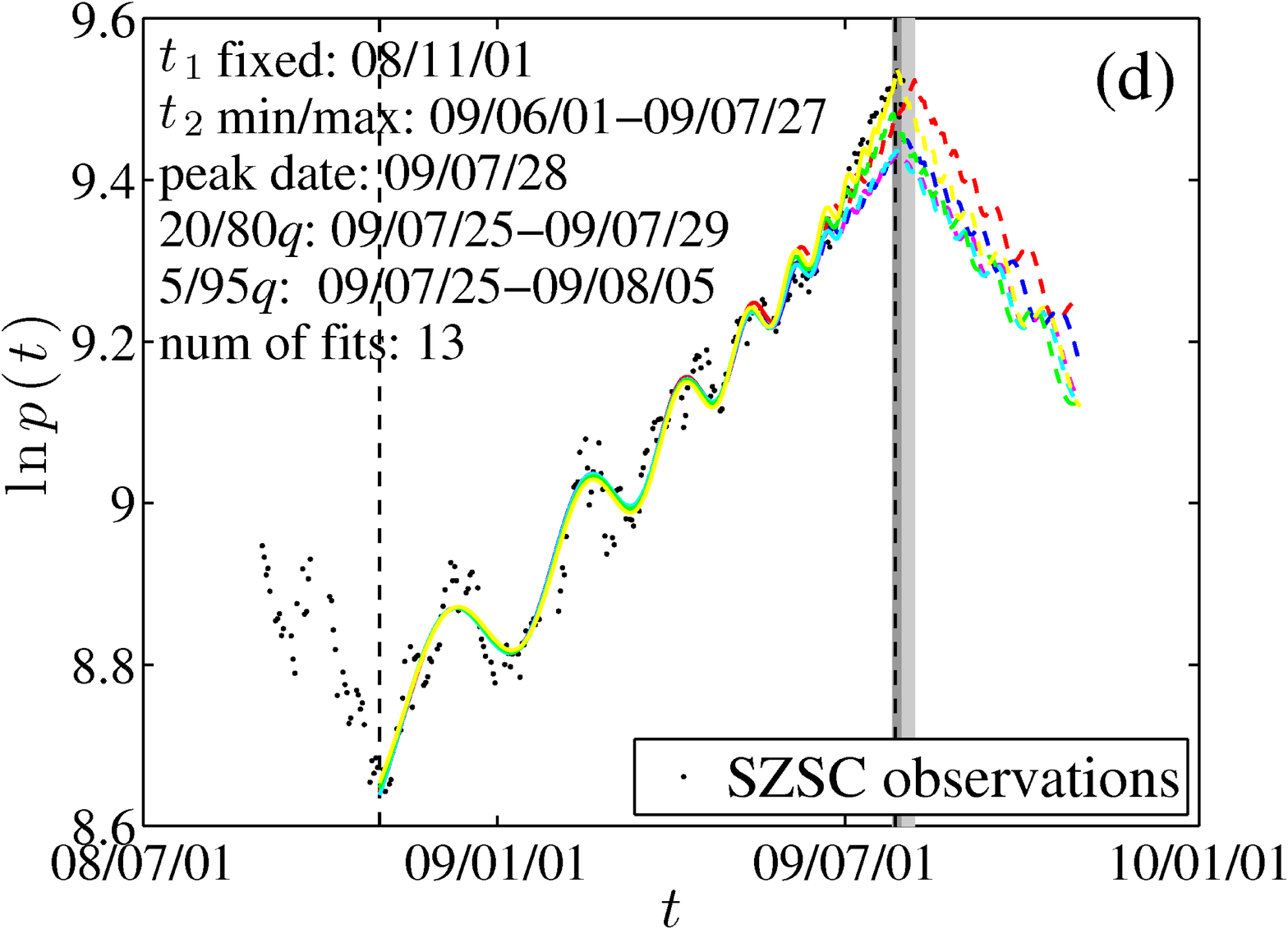}
\caption{\label{Fig:B0809} Daily trajectory of the logarithmic SSEC (a,b) and
  SZSC (c,d) index from Sep-01-2008 to Jul-31-2009 (dots) and fits to the LPPL
  formula \eqref{Eq:Landau1}. The dark and light shadow box indicate 20/80\%
  and 5/95\% quantile range of values of the crash dates for the fits,
  respectively. The two dashed lines correspond to the minimum date of $t_1$
  and the fixed date of $t_2$.  (a) Examples of fitting to shrinking windows
  with varied $t_1$ and fixed $t_2$ = Jul-31-2009 for SSEC. The six fitting
  illustrations are corresponding to $t_1$ = Oct-15-2008, Nov-07-2008,
  Dec-05-2008, Jan-05-2008, Feb-06-2008, and Feb-20-2008. (b) Examples of
  fitting to expanding windows with fixed $t_1$ = Nov-01-2008 and varied $t_2$
  for SSEC. The six fitting illustrations are associated with $t_2$ =
  Jun-01-2009, Jun-10-2009, Jun-19-2007, Jun-29-2007, Jul-13-2007,
  Jul-27-2007. (c) Examples of fitting to shrinking windows with varied $t_1$
  and fixed $t_2$ = Jul-31-2009 for SZSC. The six fitting illustrations are
  corresponding to $t_1$ = Oct-15-2008, Nov-03-2008, Nov-26-2008, Dec-19-2008,
  Jan-14-2008, and Jan-23-2008. (d) Examples of fitting to expanding windows
  with fixed $t_1$ = Dec-01-2005 and varied $t_2$ for SZSC. The six fitting
  illustrations are associated with $t_2$ = Jun-01-2009, Jun-10-2009,
  Jun-19-2007, Jun-29-2007, Jul-13-2007, Jul-27-2007.}
\end{figure}

The main results of our analysis for the 2008-2009 bubble is illustrated in
Fig.~\ref{Fig:B0809}. The SSEC and SZSC index series between Oct-15-2008 and
Jul-31-2009 have been calibrated to the LPPL formula given by
Eq.~\eqref{Eq:Landau1} in shrinking and expanding windows. The shrinking
windows are obtained by increasing the starting date $t_1$ from Oct-15-2008 to
Apr-31-2009 with a step of five days and keeping the last day $t_2$ fixed at
Jul-31-2009. The expanding windows are obtained by fixing the starting day
$t_1$ at Nov-01-2008 and moving the ending day $t_2$ away from $t_1$ from
Jun-01-2009 to Jul-31-2009 in increments of five days. The results are filtered
by the LPPL conditions, resulting in 38 (respectively 13) fits for SSEC and 28
(respectively 13) fits for SZSC in shrinking windows (respectively increasing
windows). Figures~\ref{Fig:B0809}(a) and (c) illustrate six chosen fitting
results of the shrinking windows for SSEC and SZSC and
Figures~\ref{Fig:B0809}(b) and (d) illustrate six fitting results of the
increasing time intervals for SSEC and SZSC. The dark and light shadow boxes in
the figures indicate 20\%/80\% and 5\%/95\% quantile range of values of the
crash dates for the fits that survived filtering. Our calibration confirms the
faster-than-exponential growth of the market index over this time interval, a
clear diagnostic of the presence of a bubble. It also diagnoses that the
critical time $t_c$ for the end of the bubble and the change of market regime
lies in the interval August, 1-26, 2009 for SSEC and August, 3-9, 2009 for SZSC
(20\%/80\% quantile confidence interval).

\subsubsection{Lomb analysis, parametric approach}
\label{sec:2009-results-lomb}

We use the Lomb spectral analysis technique to further investigate the
log-periodic oscillations of Eq.~\eqref{Eq:Landau1} in both indexes from
Oct-15-2008 to Jul-31-2009.  First, we calculate the detrended residuals $r(t)$
in all the surviving LPPL windows and calculate the Lomb periodogram. The
highest peak $P_N^{\max}$ and its abscissa $\omega_{\rm{Lomb}}$ are extracted
from the residual Lomb periodograms to plot as points in
Fig.~\ref{Fig:Lomb:B0809}(a). The inset plots $\omega_{\rm{fit}}$ with respect
to $\omega_{\rm{Lomb}}$. One can see that most pairs of $(\omega_{\rm{Lomb}},
\omega_{\rm{fit}})$ are overlapping on the line $y=x$ and the other pairs are
located on the line $y=2x$, confirming the existence of a strong log-periodic
component. Fig.~\ref{Fig:Lomb:B0809}(a) shows four clusters of values for
$\omega_{\rm Lomb}$ around $\omega_{\rm{Lomb}}^1 =5 \pm 1$,
$\omega_{\rm{Lomb}}^2 =10 \pm 1$, $\omega_{\rm{Lomb}}^{3-4} = 17.5 \pm 2$ and
$\omega_{\rm{Lomb}}^5 = 27 \pm 2$. The first value $\omega_{\rm{Lomb}}^1$ can
be interpreted as the main angular log-frequency, while the others are the
harmonics of order $2$ to $5$, with a rather large noise on
$\omega_{\rm{Lomb}}^{3-4}$ and $\omega_{\rm{Lomb}}^5$.

\subsubsection{Lomb analysis, non-parametric $(H, Q)$ approach}
\label{sec:2009-results-lomb-hq}

\begin{figure}[htp]
\centering
\includegraphics[width=7cm]{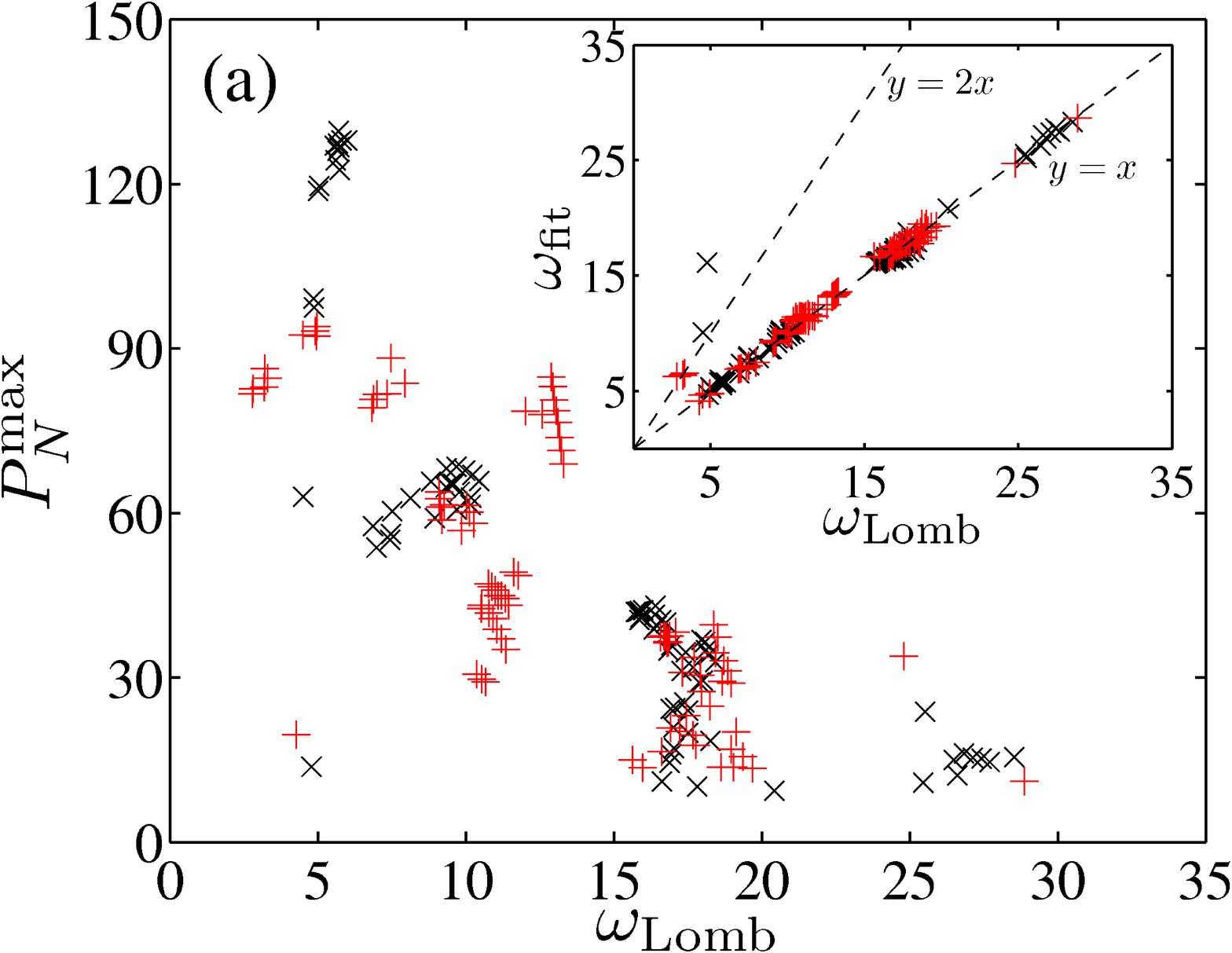}
\includegraphics[width=7cm]{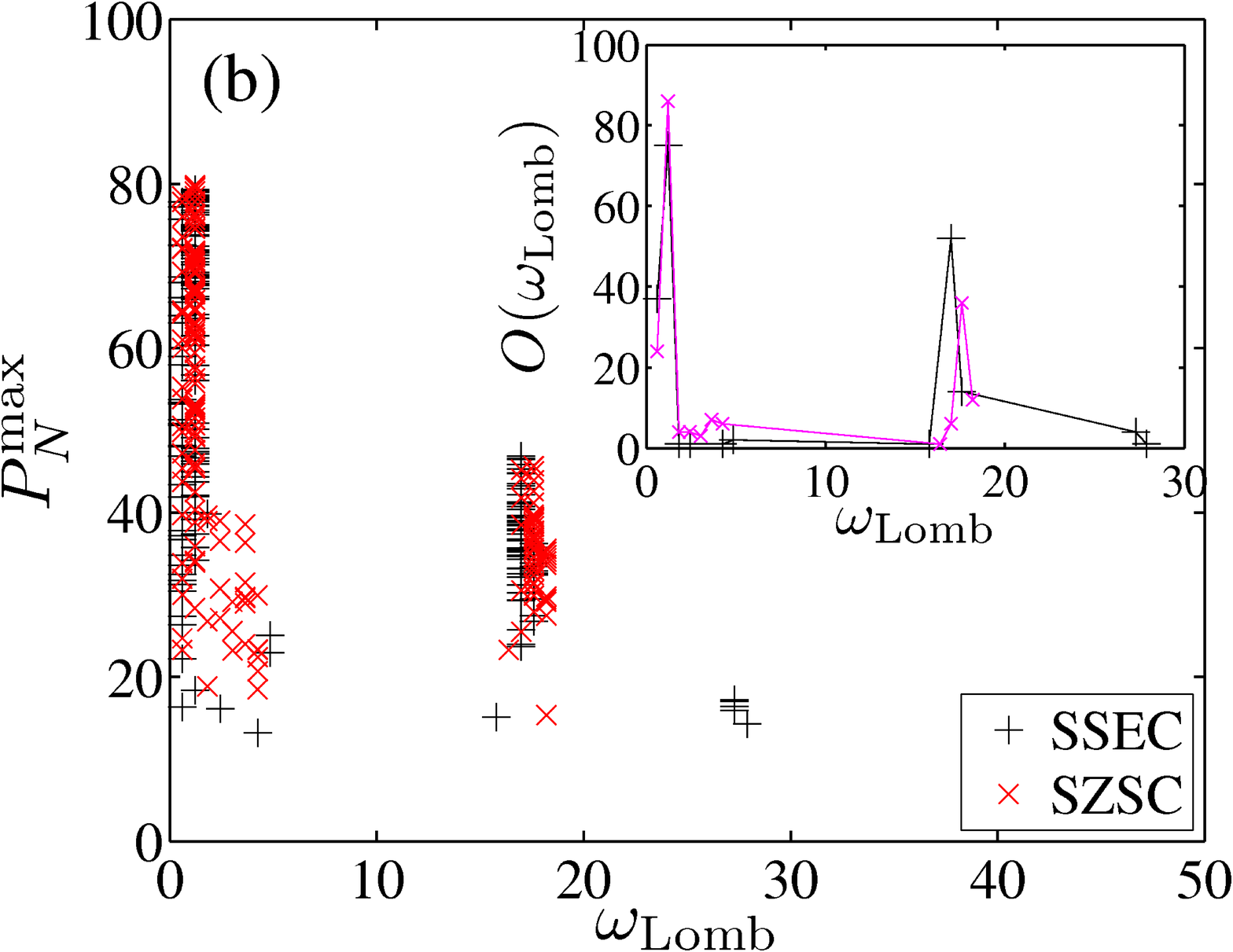}
\caption{\label{Fig:Lomb:B0809} Detection of log-periodicity in the Chinese
  bubble from 2008 to 2009. (a) Plots of $P_N^{\max}$ with respect to
  $\omega_{\rm{Lomb}}$ for different LPPL calibrating windows. The inset
  illustrates the dependence of $\omega_{\rm{fit}}$ on
  $\omega_{\rm{Lomb}}$. (b) Bivariate distribution of pairs
  $(\omega_{\rm{Lomb}},P_N^{\max})$ for different pairs of $(H,q)$ of
  $(H,q)$-derivatives $D_q^H \ln p(t)$, defined by formula (\ref{Eq:Hq}). The
  inset depicts the empirical frequency distribution of $\omega_{\rm{Lomb}}$.}
\end{figure}

We next scan a 21 $\times$ 9 rectangular grid in the $(H,q)$ plane with $H \in
[-1, 1]$ and $q \in [0.1, 0.9]$, both in steps of 0.1, using $t_c=$
Jul-31-2009, a date four days before the peak of the SSEC index.  We calculate
the corresponding $(H,q)$-derivatives $D_q^H \ln p(t)$, defined by formula
(\ref{Eq:Hq}), for this set of $H$ and $q$ values.  For each obtained $D_q^H
\ln p(t)$, we estimate the Lomb periodogram and plot the highest Lomb peak
$P_N^{\max}$ as a function of its abscissa $\omega_{\rm{Lomb}}$ in
Fig.~\ref{Fig:Lomb:B0809}(b). The inset shows the simple histogram of
$\omega_{\rm{Lomb}}$.

For the Shanghai index (SSEC), the three most prominent clusters correspond to
$\omega_{\rm{Lomb}}^0 = 1.12 \pm 0.66$, $\omega_{\rm{Lomb}}^1 = 4.0 \pm 1$ and
$\omega_{\rm{Lomb}}^3 = 17.6 \pm 2.7$.  We interpret the first cluster,
$\omega_{\rm{Lomb}}^0$, as due to the noise decorating the power-law. This is
because it corresponds to a component with less than one full period within the
interval of the $\ln(t_c -t)$ variable investigated here.  According to
extensive tests performed in synthetic time series \citep{Huang-Johansen-Lee-Saleur-Sornette-2000-JGR},
it is a spurious peak associated the most probable partial oscillations of a
noisy signal.  We identify the second cluster, $\omega_{\rm{Lomb}}^1$, as the
fundamental angular log-frequency for SSEC.  The second cluster around
$\omega_{\rm{Lomb}}^3$ is compatible with interpreting it as being the third
harmonics. It is notable that the second harmonic is not visible in this
distribution, a phenomenon which has been reported for other systems
\citep{Johansen-Sornette-Hansen-2000-PD} and can be rationalized from a renormalization group analysis
\citep{Gluzman-Sornette-2002-PRE}.

\subsection{Unit root tests of the 2005-2007 and 2008-2009
Chinese bubbles}
\label{sec:results-2009-unit-root-tests}

We apply unit root tests to the series of residuals for each $[t_1, t_2]$
interval, where the residuals are calculated by subtracting the model from the
observations.  The goal is to investigate the stationarity of the residuals to
determine if a mean-reversal Ornstein-Uhlenbeck process is a good model for
them.  The null hypothesis of the unit-root test is that the series being
tested is non-stationary.  Rejection of the null hypothesis, then, implies that
the series (or residuals in our case) is stationary.  Refer to
Section~\ref{sec:ornstein-uhlenbeck} for further details.

We calibrate the LPPL model Eq.~\eqref{Eq:Landau1} to the SSEC and SZSC indexes
in two representative intervals.  The first interval is from Dec-01-2005 to
Oct-10-2007 and the other is from Oct-15-2008 to Jul-31-2009.  We scan each
interval with growing and shrinking windows, as described above, and report the
fraction $P_{\rm{LPPL}}$ of these different windows that meet the LPPL
conditions, Eq.~\eqref{eq:LPPL-condition}.  We then calculate the conditional
probability that, out of the fraction $P_{\rm{LPPL}}$ of windows that satisfy
the LPPL condition, the null hypothesis of non-stationarity is rejected for the
residuals.  Results of these tests are shown in Table~\ref{Tb:r:OU:unitroot}.

For the time interval from Dec-01-2005 to Oct-10-2007 for the SSEC index
(respectively the SZSC index), the fraction $P_{\rm{LPPL}} = 56.9\%$
(respectively $67.6\%$) of the windows satisfy the LPPL conditions.  All of the
fitting residuals of both indexes reject the null hypothesis at significance
level 0.01 based on the two tests, implying that the residuals are
stationary. For the fraction of windows which satisfy the LPPL conditions, the
fitting residuals of the SSEC (respectively SZSC) index can be regarded as
generated by a stationary process at the $99.9\%$ (respectively $99\%$)
confidence level.

For the time interval from Oct-15-2008 to Jul-31-2009, we find that a fraction
$P_{\rm{LPPL}} = 94.4\%$ (respectively $74.7\%$) of the windows satisfy the
LPPL conditions for SSEC (respectively SZSC). All of the fitting residuals of
both indexes reject the null hypothesis at significance level 0.001 based on
the two tests, implying that the residuals are stationary. For the fraction of
windows which satisfy the LPPL conditions, the fitting residuals of both
indexes can be considered as a stationary process at the $99.9\%$ confidence
level.

\begin{table}[htp]
  \caption{\label{Tb:r:OU:unitroot} Unit-root tests on the LPPL fitting
    residuals for SSEC and SZSC index in our two calibrating ranges.
    $P_{\rm{LPPL}}$ denotes the fraction
    of windows that satisfy the LPPL condition. $P_{\rm{StationaryResi.|LPPL}}$
    denotes the conditional probability that, out of the fraction
    $P_{\rm{LPPL}}$ of windows that satisfy the LPPL condition,
    the null unit test for non-stationarity is rejected for the residuals.}
 \medskip
 \centering
 \begin{tabular}{cccclrrrr}
   \hline \hline
   \multirow{3}*[2mm]{index} & \multirow{3}{1.5cm}[2mm]{calibrating range} &
   \multirow{3}{1.5cm}[2mm]{number of windows} &
   \multirow{3}*[2mm]{$P_{\rm{LPPL}}$} & \multirow{3}{1cm}[2mm]{signif. level}
   & \multicolumn{3}{c}{percentage of rejecting $H_0$} &
   \multirow{3}*[2mm]{$P_{\rm{StationaryResi.|LPPL}}$} \\
   \cline{6-8}
   & & & & & Phillips-Perron & & Dickery-Fuller & \\
   \hline
   \multirow{3}*[2mm]{SSEC} & \multirow{3}{1.5cm}[2mm]{2005/12/01 2007/10/10} &
   \multirow{3}*[2mm]{146} & \multirow{3}*[2mm]{56.9\%} & $\alpha = 0.01$ &
   100.0\% && 100.0\% & 100.0\%\\
   \cline{5-9}
   & & & & $\alpha = 0.001$ & 95.2\% && 95.2\% & 100.0\% \\
   \hline
   \multirow{3}*[2mm]{SZSC} & \multirow{3}{1.5cm}[2mm]{2005/12/01 2007/10/10} &
   \multirow{3}*[2mm]{139} & \multirow{3}*[2mm]{67.6\%} & $\alpha = 0.01$
   &100.0\%&& 100.0\% & 100.0\%\\
   \cline{5-9}
   & & & & $\alpha = 0.001$ &81.3\%&& 81.3\% & 72.3\%\\
   \hline
   \multirow{3}*[2mm]{SSEC} & \multirow{3}{1.5cm}[2mm]{2008/10/15 2009/07/31} &
   \multirow{3}*[2mm]{54} & \multirow{3}*[2mm]{94.4\%} & $\alpha = 0.01$ &
   100.0\% && 100.0\% & 100.0\%\\
   \cline{5-9}
   & & & & $\alpha = 0.001$ & 100.0\% && 100.0\% & 100.0\% \\
   \hline
   \multirow{3}*[2mm]{SZSC} & \multirow{3}{1.5cm}[2mm]{2008/10/15 2009/07/31} &
   \multirow{3}*[2mm]{54} & \multirow{3}*[2mm]{74.7\%} & $\alpha = 0.01$
   &100.0\%&& 100.0\% & 100.0\%\\
   \cline{5-9}
   & & & & $\alpha = 0.001$ &100.0\%&& 100.0\% & 100.0\%\\
   \hline\hline
 \end{tabular}
\end{table}

\section{Prior prediction of both crashes}
\label{sec:pred-both-crash}

We make the title of this section pleonastic to emphasize that we predicted
both crashes with our techniques \textit{before} the actual dates of the
observed peaks in the two indexes.  The previous sections have presented more
thorough `post-mortem' analyses performed after the observed crashes.  This
section documents the specific but simpler predictions that we announced in
advance.

\subsection{2005-2007 bubble}
\label{sec:predict-2007-bubble}

Two of us (WXZ and DS) performed a LPPL analysis in early September 2007, which
led to (i) a diagnostic of an on-going bubble and (ii) the prediction of the
end of the bubble in early 2008. One of us (DS) communicated this prediction on
October 18, 2007 at a prominent hedge-fund conference in Stockholm. The
participants, managers of top global macro hedge-funds, constitute arguably the
best proxy for the academic idealization of ``rational investors'' with access
to almost unlimited resources and with the largest existing incentives to
motivate themselves to acquire all possible relevant information and trade
accordingly. These participants responded that the predicted change of regime
was impossible because, in their opinion, the Chinese government would prevent
any turmoil on the Chinese stock market until at least the end of the Olympic
Games in Beijing (August 2008). After the communication of October 18, 2007,
the Hang Seng China Enterprises Index (HSCEI) reached the historical high
20609.10 on 2 November 2007.  Afterwards, the first valley HSCEI=15460.72
(-25\% from historical high) was reached on 22 Nov 2007 and the bottom
HSCEI=4792.37 (-77\% from historical high) was on 29 Oct 2008. On 19 March
2008, HSCEI=11379.91 was another deep valley.  These drops occurred after a
six-fold appreciation of the Chinese market from mid-2005 to October 2007.

\subsection{2008-2009 bubble}
\label{sec:predict-2009-bubble}

\subsubsection{The announcement of the prediction}

On 10 July 2009, we submitted our prediction online to the \url{arXiv.org}
\citep{Bastiaensen-Cauwels-Sornette-Woodard-Zhou-2009-XXX}, in which we gave the 20\%/80\% (respectively
10\%/90\%) quantiles of the projected crash dates to be July 17-27, 2009
(respectively July 10 - August 10, 2009).  This corresponds to a $60\%$
(respectively $80\%$) probability that the end of the bubble occurs and that
the change of regime starts in the interval July 17-27, 2009 (respectively July
10 - August 10, 2009).  Redoing the analysis 5 days later with $t_2 =$ July 14,
2009, the predictions tightened up with a 80\% probability for the change of
regime to start between July 19 and August 3, 2009 (unpublished).

\begin{figure}[htp]
\centering
\includegraphics[width=10cm]{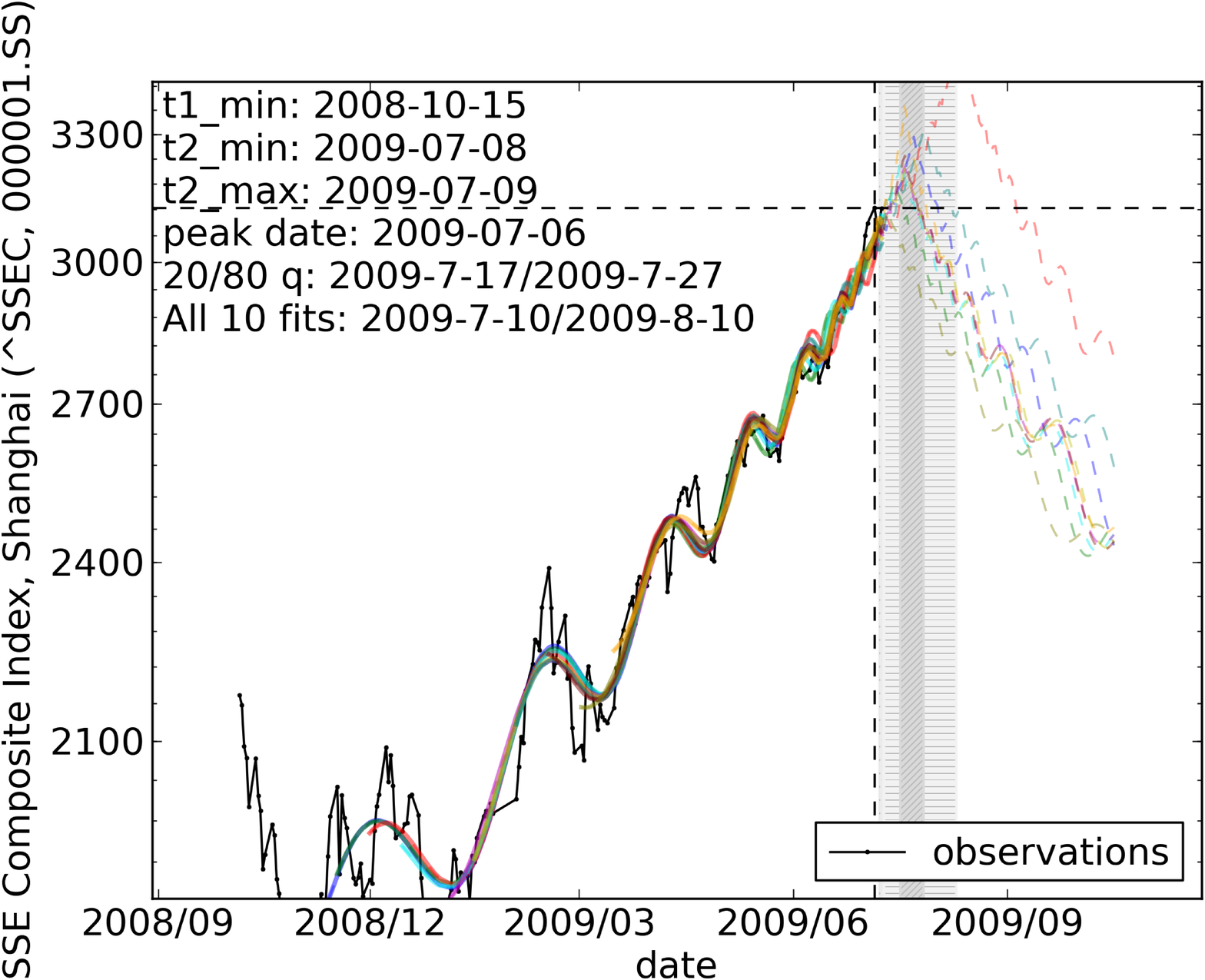}
\caption{\label{Fig:SSEC_2009-bubble} Shanghai Composite Index with LPPL
  result, as presented in the July 10, 2009 \protect\url{arXiv.org} submission
  of \citet{Bastiaensen-Cauwels-Sornette-Woodard-Zhou-2009-XXX}.  }
\end{figure}

The following paragraph and figure \ref{Fig:SSEC_2009-bubble} are reproduced
from the online (and un-refereed) prediction, which is available in its initial
form at the corresponding URL \citep{Bastiaensen-Cauwels-Sornette-Woodard-Zhou-2009-XXX}.

\begin{quotation}
  The result of the analysis is summarized below in the Figure. We analyzed the
  Shanghai SSE Composite Index time series between October 15, 2008 and July 9,
  2009. We increased the starting date of the LPPL analysis in steps of 15 days
  while keeping the ending date fixed, resulting in 10 fits. The figure shows
  observations of the SSEC Index as black dots (joined by straight lines) and
  the LPPL fits as smooth lines until the last day of analysis. The y-axis is
  logarithmically scaled, so that an exponential function would appear as a
  straight line and a power law function with a finite-time singularity would
  appear with a slightly upward curvature. Note that the LPPL fits to the
  observations exhibit this slightly upward curvature. The vertical and
  horizontal dashed lines indicate the date and price of the highest price
  observed, July 6, 2009. Extrapolations of the fits to 100 days beyond July 9,
  2009 are shown as lighter dashed lines. The darker shaded box with diagonal
  hatching indicates the 20\%/80\% quantiles of the projected crash dates, July
  17-27, 2009. The lighter shaded box with horizontal hatching indicates the
  range of all 10 projected crash dates, July 10 - August 10, 2009. These two
  shaded boxes indicate the most probable times (with the associated confidence
  levels) to expect peak and possible subsequent crash of the Index. The
  parameters of the fit confirm the faster-than- exponential growth of the
  Shanghai SSE Composite Index over this time interval, a clear diagnostic of
  the presence of a bubble.
\end{quotation}

\subsubsection{What actually happened}

\begin{figure}[htp]
\centering
\includegraphics[width=10cm]{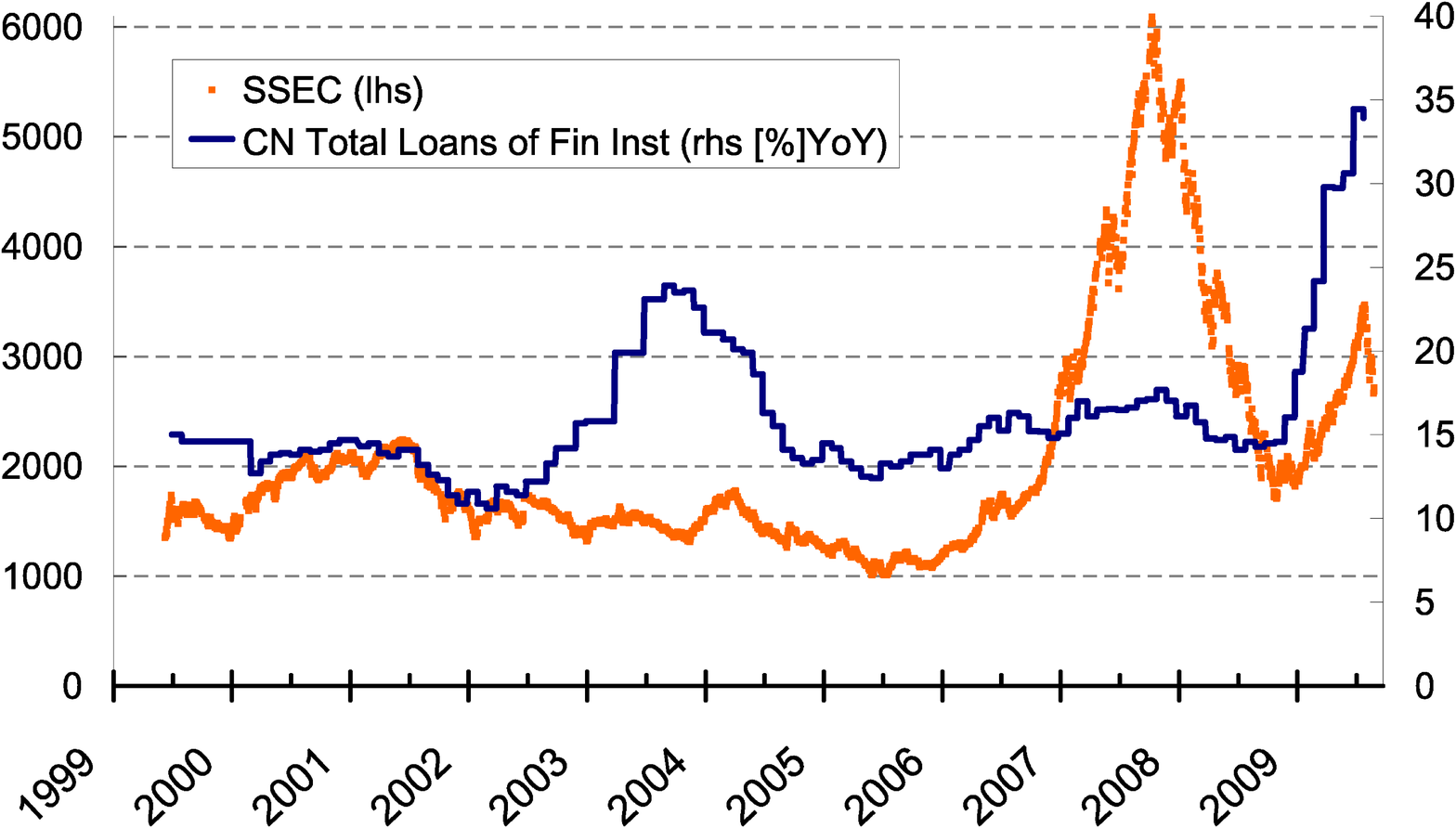}
\caption{\label{Fig:shcomp_vs_loans} (left axis, dots) SSEC compared with Total
  Loans of Financial Institutions as reported by The People's Bank of China                                                                                    (``Summary of Sources \& Uses of Funds of Financial Institutions'' http://www.pbc.gov.cn/english)
 (right axis, solid line) YoY \% monthly
  change. This shows graphically the widespread belief that the credit growth
  has fueled the last Chinese equity bubble.  }
\end{figure}

On July 29, 2009, Chinese stocks suffered their steepest drop since November
2008, with an intraday bottom of more than $8\%$ and an open-to-close loss of
more than $5\%$.  The market rebounded with a peak on August 4, 2009 before
plummeting the following weeks. The SSEC slumped 22 percent in August, the
biggest decline among 89 benchmark indices tracked world wide by Bloomberg, in
stark contrast with being the no. 1 performing index during the first half of
this year. These striking facts show the detachment of the Chinese equity
market from other markets.  This bubble was probably nucleated by China's
central government's reaction to the global financial crisis. Besides
announcing the huge stimulus plan on 9 November 2008, a loose monetary policy
and regulations caused massive new loan issuance as shown in
Fig.~\ref{Fig:shcomp_vs_loans}. With overproduction and lower global demands,
analysts estimate that up to 50\% of the increase in credit was used to
speculate in equities, property and commodities \footnote{see, e.g., BNP Paribas FX Weekly Strategist: China Lending Support (31 July 2009); RBS, Local Markets Asia, Alert China: A savings glut is causing problems (3 July 2009)}. Rumours of asset bubbles were widely
heard in the market, but when or if they might crash was unknown as usual.

Note that the change of regime in the SSEC occurred while the total loan of
financial institutions was still growing at close to its peak YoY 35\% monthly
rate. This illustrates that the change of regime has occurred in absence of any
significant modification of the economic and financial conditions or any
visible driving force. This observation, which should be surprising to most
economists and analysts, is fully expected from the mathematical and
statistical physics of bifurcations and phase transitions on which our LPPL
methodology is based: a possibly vanishingly small change of some control
parameter may lead to a macroscopic bifurcation or phase transition.  Rather
than leading to an absence of predictability, the accelerating susceptibility
of the system associated with the approach towards the critical point can be
diagnosed, as we have shown.  The very clear change of regime documented here
provides a case-in-point demonstrating this concept of an emergent rupture
point characterizing the end of the bubble.

\begin{figure}[htp]
\centering
\includegraphics[width=10cm]{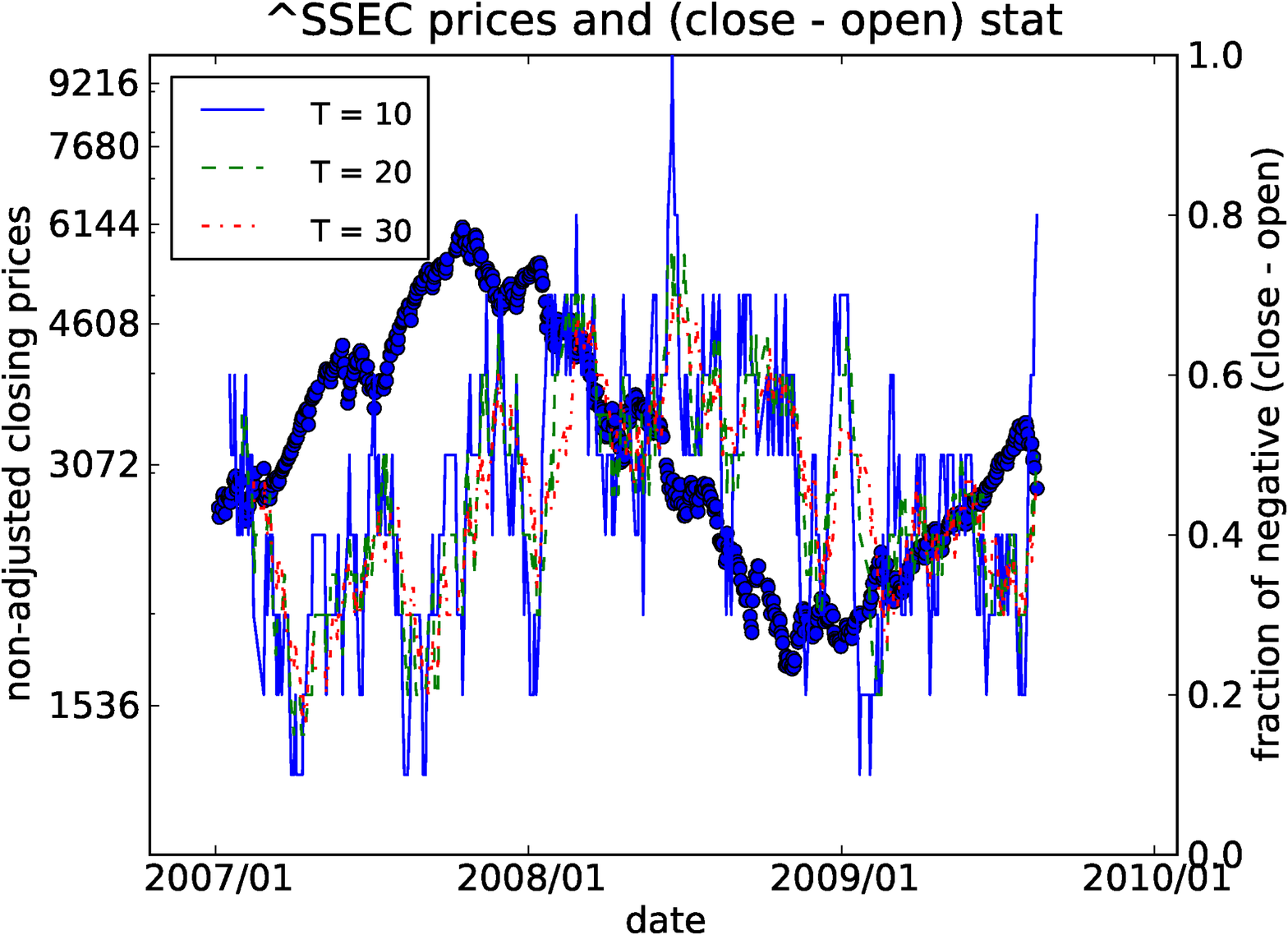}
\caption{\label{Fig:ssec_close_open_stat} Left scale: SSE Composite index from
  Jan. 2007 to August 25, 2009 (closing price 2980.10).  Right scale: fraction
  of days with negative (close-open) in moving windows of length $T=10$ days
  (continuous blue line), $T=20$ days (dashed green line) and $T=30$ days
  (dotted-dashed red line).  }
\end{figure}

Figure \ref{Fig:ssec_close_open_stat} presents the evolution with time of the
close-open statistic introduced in subsection \ref{y3ykbwgw} over the period
from Jan. 2007 to August 25, 2009.  The low (respectively high) values of the
index correlate well with the ascending (respectively descending) trend of the
market. One can also observe the recent remarkably abrupt jump upward of the
close-open statistic at the time scale $T=10$ days, confirming the existence of
a sudden change of regime.

These events unfolded in a rather bullish atmosphere for the Chinese stock
markets.  For instance, Bloomberg reported on July 30, 2009 that billionaire
investor Kenneth Fisher emphasized the great success of China's economy
compared to the rest of the World and that speculation that the ``Chinese
government will limit bank loans is unfounded.''  Anecdotal sampling of
comments on Chinese online forums suggests a majority of doubters until August
12, after which a majority endorsed the notion of a change of regime.  Several
commentators stressed again that predictions of Chinese stock markets cannot be
correct since China's stock markets are heavily influenced by policies (known
as a policy market).  These comments are similar to the disbelief of hedge-fund
managers mentioned in subsection \ref{sec:predict-2007-bubble} concerning the
prediction of the change of regime at the end of 2007, before the 2008 Beijing
Olympics.

\section{Discussion}
\label{Sec:LPPL:Conclusion}

We have performed a detailed analysis of two financial bubbles in the Chinese
stock markets by calibrating the LPPL formula \eqref{Eq:Landau1} to two
important Chinese stock indexes, Shanghai (SSEC) and Shenzhen (SZSC) from May
2005 to July 2009. Bubbles with the property of faster-than-exponential price
increase decorated by logarithmic oscillations are observed in two distinct
time intervals within the period of investigation for both indexes. The first
bubble formed in the middle of 2005 and burst in October 2007.  The other
bubble began in November 2008 and reached a peak in early August 2009.

Our back tests of both bubbles find that the LPPL model describes well the behavior of
faster-than-exponential increase corrected by logarithmic oscillations in both
market indexes. The evidence for the presence of
log-periodicity is provided by applying Lomb spectral analysis on the detrended
residuals and $(H,q)$-derivative of market indexes. Unit-root tests, including
the Phillips-Perron test and the Dickey-Fuller test, on the LPPL fitting
residuals confirm the O-U property and, thus, stationarity in the residuals,
which is in good agreement with the consistent model of `explosive' financial
bubbles \citep{Lin-Ren-Sornette-2009-XXX}.

While the present paper presents post-mortem analyses,
we emphasize that we predicted the presence and expected critical date $t_c$ of
both bubbles in advance of their demise \citep{Sornette-2007,Bastiaensen-Cauwels-Sornette-Woodard-Zhou-2009-XXX}. These two successes
prolong the series of favorable outcomes following the prediction
of the peak in mid-2004 of the real-estate bubble in the UK by two of us
\citep{Zhou-Sornette-2003a-PA}, of the peak in mid-2006 of the US housing
bubble by two of us \citep{Zhou-Sornette-2006b-PA} and of the peak in July 2008 of the global
oil bubble by three of us \citep{Sornette-Woodard-Zhou-2009-PA}.

But not all predictions based on the present methodology have fared so well.
In particular, \citet{Lux-2009} and \citet{Rosser-2008-ACS} have
raised severe objections, following the failure of the well-publicized prediction
published in 2002 that the U.S. stock market would follow a downward
log-periodic pattern \citep{Sornette-Zhou-2002-QF}. How can one make
sense of these contradictory claims? We summarize the present state of the art
as follows.
\begin{enumerate}
\item {\bf Prediction of the end of bubbles should not be confused with
predictions based on extrapolation, such as those associated with antibubbles}.
There is a confusion between predicting crashes, on the one hand,
and predicting the continuation of an ``antibubble'' bearish regime, on the other hand.
It seems that both \citet{Lux-2009} and \citet{Rosser-2008-ACS} amalgamate
these two issues, when they focus on the failure
of the antibubble prediction in \citet{Sornette-Zhou-2002-QF} and conclude that ``Sornette and his collaborators
failed to forecast future crashes.''

There is indeed a fundamental difference between, on the one hand, (i) the prediction of the
{\it end} of a bubble analyzed here, which is characterized by its critical time $t_c$ and, on the other hand,
(ii) the extrapolation of an ``antibubble'' pattern. This difference is similar to
that between (i) the prediction of the approximate
parturition time of a foetus
on the basis of the recording of key variables obtained during its maturation
in the uterus of his mother
and (ii) the prediction of the death of this individual later in old age from an extrapolation of medical
variables recorded during his adult life. The former  (i)
is associated with the maturation phase (the financial bubble versus the uterus-foetus
development). It has a rather well-defined critical time which signals
the transition to a new regime (crash/stagnation/bearish versus birth).
In contrast, the latter (ii) may be influenced by a variety of factors, particularly exogenous,
which may shorten or lengthen the life of the antibubble or of the individual.
One should thus separate the
statistics of successes and failures in the prediction of bubbles on the one hand
and in the prediction of continuation of LPPL antibubble patterns on the other hand.
This was the spirit of the experiment proposed by \citet{Sornette-Zhou-2002-QF}
to test the {\it distinct} hypothesis that bearish regimes following a market peak
could be predicted when associated with LPPL patterns.

\item  {\bf Intrinsic limits of the prediction of the end of an antibubble}.
For the reasons just mentioned, it is still an open problem to
determine when an antibubble ends. Our methods show that one cannot avoid
a delay of about 6 months before identifying the end of an antibubble  \citep{Zhou-Sornette-2005-PA}.
This is a partial explanation for the failure of the 2002  antibubble prediction  \citep{Sornette-Zhou-2002-QF}.

\item {\bf Track record of the antibubble method}.
However, one should not forget that, taken as a distinct class separated from that
of diagnosing bubbles and their ends,  the predictions based on the antibubble method can count several
past successes: (a) on  the Nikkei antibubble \citep{Johansen-Sornette-1999-IJMPC,Johansen-Sornette-2000-IJMPC}
and (b) on the Chinese stock market antibubble \citep{Zhou-Sornette-2004a-PA}, in
addition to the failure mentioned above. This track record is insufficient to
conclude. More tests in real time should be performed and rigorous methods developed
to assess the statistical significance of short catalogs of success/failure predictions
can be applied, based on ``roulette'' approaches (see Chapter 9 in \citet{Sornette-2003}, Bayes' theorem
\citep{Johansen-Sornette-2000-IJMPC} and Neyman-Pearson or error diagrams \citep{Molchan-1990-PEPI,Molchan-1997-PAG}.

\item {\bf What we learned from the antibubble prediction failure}.
\citet{Lux-2009} and \citet{Rosser-2008-ACS} are right to
stress that the 2002 antibubble prediction of \citet{Sornette-Zhou-2002-QF}
failed. However, a post-mortem analysis in \citet{Zhou-Sornette-2005-PA}
has revealed an interesting fact.
While the prediction failed when the S\&P500 is valued in U.S. dollars,
it becomes quite accurate when expressed in euro or British pounds
\citep{Zhou-Sornette-2005-PA}. A plausible interpretation would be that
the energetic Fed monetary policy of decreasing its lead rate from 6.5\% in 2000
to 1\% in 2003 has boosted the stock market in local currency from 2003 on,
but has degraded the dollar, so that the net effect
was that the value of the US stock market from
an international reference point was unfolding as expected from the analysis
of \citet{Sornette-Zhou-2002-QF}.
We do not claim that this changed the failure into a success. Instead, it illustrates
the effect of monetary feedbacks that have to be included in improved models
incorporating fundamental factors, for instance in the spirit of \citet{Zhou-Sornette-2006a-PA}.

\item {\bf Track record for diagnosing bubbles and their ends}.
Our group has announced {\it advanced} prediction (not just in retrospect) of bubbles
and their end (often a crash). The status of these predictions as of 2002 has been discussed
in details in Chapter 9 of \citet{Sornette-2003}'s book.
As mentioned above, subsequent successes include the predictions
of the peak in mid-2004 of the real-estate bubble in the UK by 
\citet{Zhou-Sornette-2003a-PA}, of the peak in mid-2006 of the US housing
bubble by \citet{Zhou-Sornette-2006b-PA} and of the peak in July 2008 of the global
oil bubble by \citet{Sornette-Woodard-Zhou-2009-PA}.
The present analysis on two bubbles in the Chinese market provide additional
evidence for the relevance of LPPL patterns in the diagnostic of bubbles.

\item {\bf Recent improvements in methodology}.
While the core model (or forecasting system) has not changed much since
the late 1990s, several new developments used here allows us to quantify
more accurately both the reliability and the uncertainties. These
improvements include multi-window analysis, probability estimates,
and a consistent LPPL rational expectation model with mean-reverting residuals.
Note that the more recent bubble in
the Chinese indexes was detected and its end or change of regime was
predicted independently by two groups (the first four authors
from academia on the
one hand and the last two authors from industry on the other hand)
with similar results, showing that the
model has been well-documented and can be replicated by industrial
practitioners.  In addition, we stress that the method relies essentially on the
competition between the positive feedback loop of higher return anticipations
competing with negative feedback spirals of crash expectations \citep{Ide-Sornette-2002-PA}, which
is at the origin of the acceleration oscillations. In the spirit of
\citet{Lux-Marchesi-1999-Nature} and of \citet{Gallegati-Palestrini-RosserJr-2008-XXX},
this is accounted for in heterogeneous agent models
by including nonlinear fundamental investment styles competing
with nonlinear momentum trading styles \citep{Ide-Sornette-2002-PA}.
The initial JLS model of \citet{Johansen-Sornette-Ledoit-1999-JR,Johansen-Ledoit-Sornette-2000-IJTAF}
was based on a conventional neoclassical model assuming a homogeneous
rational agent, but it also enriched this set-up by introducing
heterogeneous noise traders driving a crash hazard rate. More recently, \citet{Lin-Ren-Sornette-2009-XXX} have considered an alternative framework which extends the JSL model to account for behavioral herding by using
a behavioral stochastic discount factor approach, with self-consistent
mean-reversal residuals.

\end{enumerate}

In conclusion, given all the above,
we feel this technique is the basis of a prediction
platform, which we are actively developing, motivated by the conviction that
this is the only way to make scientific progress in this delicate and crucial
domain of great societal importance, as illustrated by the 2007-2009 financial
and economic crisis \citep{Sornette-Woodard-2009-XXX}.

\bigskip {\textbf{Acknowledgments:}} The authors would like to thank Li Lin and
Liang Guo for useful discussions. This work was partially supported by the
Shanghai Educational Development Foundation (2008SG29) and the Chinese Program
for New Century Excellent Talents in University (NCET-07-0288). We also
acknowledge financial support from the ETH Competence Center Coping with Crises
in Complex Socio-Economic Systems (CCSS) through ETH Research Grant
CH1-01-08-2.  Much appreciation goes to Prof. Jan Ryckebusch of the University
of Ghent, Subatomic Physics Department, for useful discussions.

\bibliographystyle{elsarticle-harv}
\bibliography{Bibliography}

\end{document}